\documentclass[preprint]{aastex}
\usepackage{graphicx}
\usepackage{hyperref}
\usepackage{color}

\begin{document}
\title{Evidence for Photoionization Driven Broad Absorption Line Variability}
\author{Tinggui Wang\altaffilmark{1}, Chenwei Yang\altaffilmark{1}, 
Huiyuan Wang\altaffilmark{1} \and Gary Ferland\altaffilmark{2,3} 
}
\altaffiltext{1}{
CAS Key Laboratory for Research in Galaxies and Cosmology, Department of 
Astronomy, University of Science and Technology of China, Hefei, Anhui 
230026; twang@ustc.edu.cn}
\altaffiltext{2}{Department of Physics and Astronomy, University of Kentucky, Lexington, KY 40506, USA}
\altaffiltext{3}{Centre for Theoretical Atomic, Molecular and Optical Physics, School of Mathematics and Physics, Queens University Belfast, Belfast BT7 1NN, UK}

\begin{abstract}
We present a qualitative analysis of the variability of quasar broad absorption 
lines using the large multi-epoch spectroscopic dataset of the Sloan Digital Sky 
Survey Data Release 10. We confirm that variations of absorption lines 
are highly coordinated among different components of the same ion or the same 
absorption component of different ions for C~{\sc iv}, Si~{\sc iv} and N~{\sc v}. 
Furthermore, we show that the equivalent widths of the lines decrease or increase 
statistically when the continuum brightens or dims. This is further supported by the 
synchronized variations of emission and absorption line 
equivalent width, when the well established intrinsic Baldwin effect for emission 
lines is taken into account.  We find that the emergence of an absorption component 
is usually accompanying with dimming of the continuum while the disappearance of an 
absorption line component with brightening of the continuum. This suggests that the 
emergence or disappearance of a C~{\sc iv} absorption component is only the 
extreme case, when the ionic column density is very sensitive to continuum 
variations or the continuum variability amplitude is larger. These results support 
the idea that absorption line variability is driven mainly by changes in the 
gas ionization in response to continuum variations, that the line-absorbing gas 
is highly ionized, and in some 
extreme cases, too highly ionized to be detected in UV absorption lines. Due to 
uncertainties in the spectroscopic flux calibration, we cannot quantify the fraction 
of quasars with asynchronized continuum and absorption line variations.  
\end{abstract}


\section{Introduction}
It is generally accepted that quasar feedback plays a crucial role in the context 
of galaxy formation and evolution. The bimodal color distributions of galaxies 
at the present universe, and the deficiency of massive galaxies in the mass 
function of galaxies in comparison with dark-matter halo mass distributions, require 
quenching of massive galaxies (Silk 2011). The correlations between masses of supermassive 
black holes (SMBHs) and the bulge of their host galaxies suggest that the growth of 
galaxies and supermassive black holes at their centers are closely 
connected (e.g., Kormendy \& Ho 2013). Considering the high radiation efficiency 
of black hole accretion, even if a small fraction of the enormous power 
from accreting SMBH during the quasar phase goes into the interstellar medium in 
the host galaxy, it can interrupt the growth  of both the black hole and 
their host galaxy. Quasar outflow is a natural form of such feedback, given 
its ubiquity among quasars, with a potentially large kinetic power, as 
demonstrated in a few quasars with UV broad absorption lines from S {\sc iv}, 
O {\sc iv} or Fe {\sc ii} excited states or ionized X-ray absorption lines 
(e.g., Korista et al. 2008; Arav et al. 2013; Borguet et al. 2013; 
Chartas et al. 2009; Tombesi et al. 2015). Outflows are manifested as 
blueshifted broad emission lines (BELs) or broad absorption lines (BALs), 
imprinted on UV or X-ray spectra of quasars. However, the total kinetic 
energy of outflows on these scales, as well as the way they interact with 
the ISM, remains very uncertain,  because of the uncertainty in the geometry 
and the total column density of quasar absorbers, together with the 
complication of more than one stable phase (Arav et al. 2013). 

Quasar absorption lines were divided into two classes according to their origins. 
The intervening absorption lines, which are produced by galaxies and intergalactic 
material between the quasar and the observer, are not the subject of this study.  
Intrinsic absorption 
lines, formed by gas either directly associated with the quasar or in its host 
galaxy, are our main focus. Most intrinsic absorption lines are blueshifted with 
respect to the corresponding emission lines, implying that the gas is flowing out 
of the center. Intrinsic absorption lines are further classified into narrow and 
broad absorption lines (NALs and BALs) according to the width of the absorption 
line (e.g., Weymann et al. 1991; Barlow \& Sargent 1997; Barlow et al. 1997). 
In this paper, we will focus on broad absorption lines, but we do not distinguish 
BALs and mini-BALs with a width 500-2000~km~s$^{-1}$ (Hamann \& Sabra 2004) as there is a continuous 
distribution of absorption index between the two classes (e.g., Trump et al. 2006; 
Ganguly \& Brotherton 2008; Gibson et al. 2009) and there are instances where 
a source can transition from mini-BALs to BALs (e.g., Rodriguez Hidalgo et al. 2013). 
It has long been known that broad absorption lines are variable on time scales from 
years to months (Foltz et al. 1987; Turnshek et al. 1988; Smith \& Penston 1988; 
Barlow et al. 1989, 1992; Barlow 1993), but systematic studies have only recently
been carried out (Lundgren et al. 2007; Gibson et al. 2008, 2010; Filiz Ak et al. 2012, 2013; 
Capellupo et al. 2011, 2012, 2014; Wildry, Goad \& Allen 2013; Welling et al. 2014). 

Capellupo et al. (2011, 2014) monitored a sample of 24 BAL QSOs at
$z\sim 2$ over time scales of 0.02 to 8.7 years \footnote{Time will be referred in 
rest frame of quasars unless explicitly state otherwise.} and found that 65\% of 
their quasars displayed BAL variability on over several years and 39\% on 
time scales of less than one year. Similarly, Wildry, Gaod \& Allen (2014) analyzed 
2-epoch spectra of 50 BAL quasars spanning 10 months to 3.7 years, and found that 
50-60\% BALs are variable on such time scales. Using multi-epoch spectra from the 
SDSS I/II/III, Filiz Ak et al. (2012, 2013) made a statistical study of the 
largest sample consisting of 582 BAL quasars. They showed that the fraction 
of variable BALs increases with increasing observing intervals. All these studies 
also found that different 
BAL components of C~{\sc iv} in one object or the same BAL troughs of C~{\sc iv} 
and Si~{\sc iv} BALs 
vary in a coordinated manner (strengthening or weakening). These coordinated 
variations can most simply be interpreted as caused by the variation in the 
ionizing continuum (Hamann et al. 2011). In contrast, Wildy et al (2014) claimed 
that BAL variations were not 
correlated with continuum variations, casting doubt on this interpretation. In 
addition, Capellupo et al. (2014) found variable P~V absorption lines 
associated with non-black C~{\sc iv} absorption lines, which they interpreted as 
ionized gas moving across the continuum source. There is also an argument that the ionizing 
continuum may vary in a different way from the observed UV continuum due to changes in 
a presumed high-ionization shielding gas (Filiz Ak et al. 2012).  That intervening gas 
filters hard ionizing photons to prevent gas over-ionized outer layers (Murray et al. 
1995).  Also, in some cases, a new component appears or a previously observed trough 
disappears, which is interpreted as gas moving into or out of our line of sight (Hamann 
et al. 2008; Leighly et al. 2009; Krongold et al. 2010; Filiz Ak et al. 2012).   

While the cause of this variability is still a matter of debate, there is no doubt 
that absorption line variability can put useful constraints on the physical state 
and/or kinematics of the absorbing gas. In the case of photoionization, the 
variability time scale will set an upper limit on the ionization or recombination 
time scale, which depends solely on the flux density of the incident ionizing continuum 
or on the gas density.  In the case of gas moving in and out of the line of sight, 
important constraints on the gas transverse velocity and the size of clump can be 
set. In combination with specific kinematic models, this can translate into constraints 
on the distance of the absorber to the central black holes (e.g., Hall et al. 2011).

In this paper, we will examine the variability of BALs and mini-BALs and their 
correlations with continuum and emission line variability to constrain 
the physical process that leads to such variability, and to interpret the observed 
variability in the context of such physical process. The paper is organized as 
follows. We describe the sample and data analysis methods in \S 2 and 3. A 
correlation analysis is presented in \S 4. We discuss the implications and 
photoionization models in \S 5. A summary is given in \S 5.

\section{Variable Absorption Line Quasar Sample\label{dataanalysis}}

We searched the SDSS DR10 archive for quasars that were observed two or more times. 
We merged the quasar catalog of SDSS data release 7 (DR7, Schneider et al. 2010) 
with that of DR10 (P\^aris et al. 2014). Duplicated entries are removed. We compared 
this catalog with the SDSS spectroscopic catalog and selected quasars with 
multi-spectroscopic observations. To investigate the variability of C~{\sc iv}, 
Si~{\sc iv} and N~{\sc v}, we adopt a redshift cut $2.2<z\leqslant4.7$. 
To ensure detection of major absorption lines, we only keep quasars with 
at least one spectrum with signal to noise $ SNR >10$. After these screenings, we 
obtain a sample of 6250 quasars with two spectra or more. In the following subsection 
we describe the method of construction of a variable absorption line quasar sample.

\subsection{Unabsorbed Quasar Templates and Identification of Absorption Lines}

\begin{figure}
\plotone{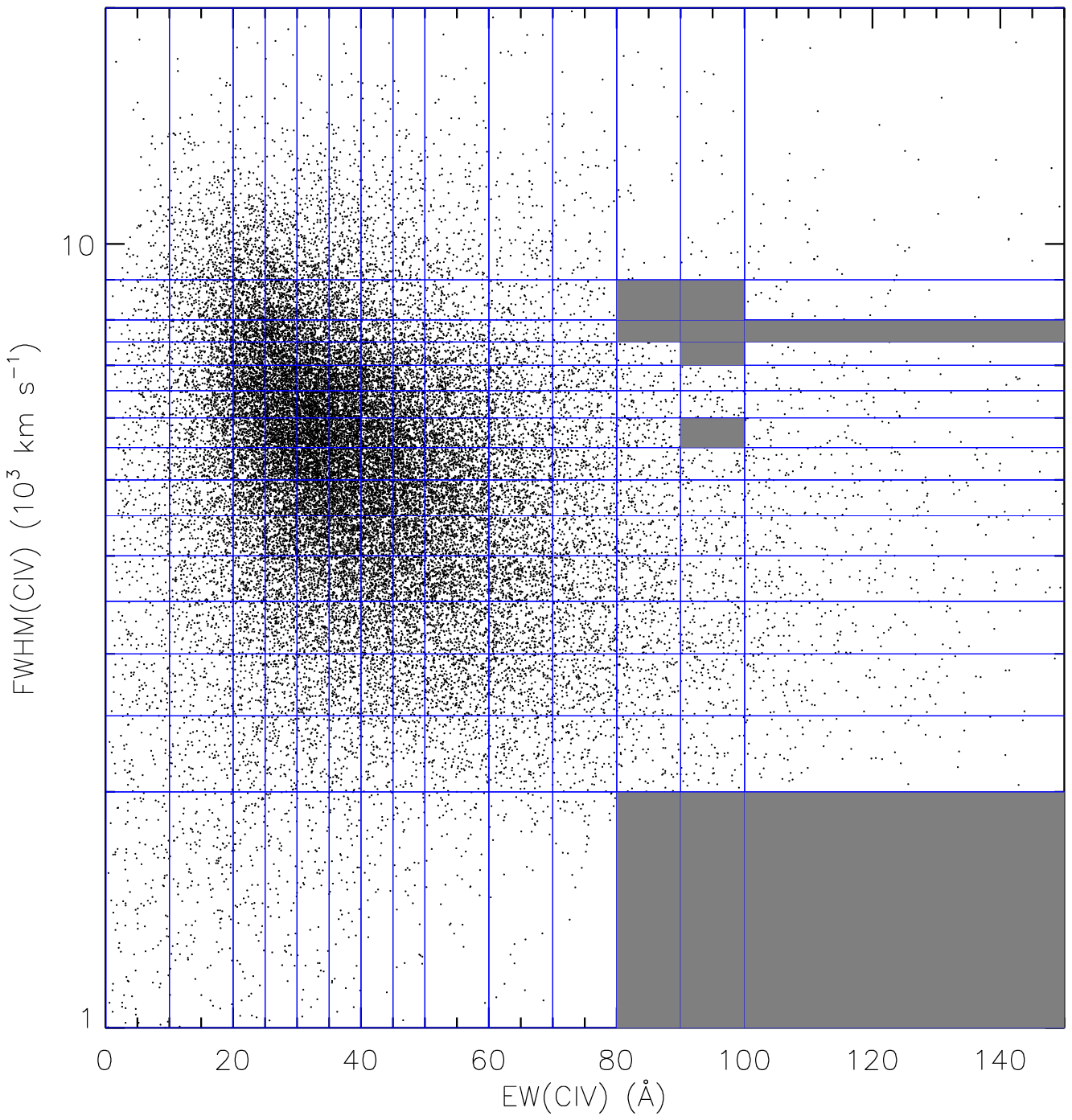}
\caption{The grids on the plane of $EW(C~{\sc iv})$ .vs. $FWHM(C~{\sc iv})$ that unabsorbed 
composite spectra are made. The scattered dots represent the data of quasars 
in SDSS DR7. No composite spectrum is made for grey area because of less than 
20 quasars in each grid.\label{comp}}  
\end{figure}

In order to identify absorption lines, we need a set of unabsorbed template 
quasar spectra. In the literature, there are several ways of constructing such 
templates: using quasar composite spectrum as an approximate (e.g., Weymann et 
al. 1991); using the reddened power-law continuum plus Gaussian or Lorentzian emission-line 
models (e.g., Filiz Ak et al. 2012), using a set of PCA spectra of quasars (e.g., 
Wildry et al. 2014), and finally adopting the best-matching unabsorbed quasars (Zhang et al. 
2014; Liu et al. 2015). In this paper, we will use a set of composite spectra with 
different properties of emission lines to account for the diversity of quasar spectra. 
To first order, a quasar emission line can be described by its relative strength (EW), 
its blueshift, and its width (FWHM). Previous studies have shown that those line parameters for 
C~{\sc iv} are correlated with each other (e.g., Corbin \& Boroson 1996; Richards et al. 
2011; Wang et al. 2011) and also with relative strength of other lines (Wang et al. 2012). 
Here, we divide 38,377 non-BAL quasars 
with $1.5<z\leq 4.0$ and $SNR_{1350}>10$  in  Data Release 7 (DR7) into different bins according to 
their line width ($FWHM$) and equivalent width (EW) of C~{\sc iv} as measured by Shen et al. 
(2011). We then make a composite spectrum for each bin. We split the $EW$(C {\sc iv}) 
and $FWHM$ plane into 210 unequal grids as shown in Figure \ref{comp}.
To ensure a sufficient number ($>$20) of spectra per bin, we use a larger grid in the 
low-density zone of  parameter space. Among the 210 grids, 10  contain less 
than 20 spectra each, so no composite spectrum is made for these grids. In the process, we 
mask bad pixels using the SDSS mask array and also mask strong narrow absorption lines in individual 
spectra. We smooth the SDSS spectrum using a Savitzky-Golay filter with 33 pixels, 4th 
degree polynomial. Any region 10\% lower than the smoothed 
spectrum and with a width of at least 3 pixels is considered as a narrow absorption line. 
We replace the flux of these pixels with the smoothed ones. Finally we 
check the templates by eye. This yields a total of 200 templates.  With this number 
of templates we are able to fit emission lines and continuum in individual objects.

\begin{figure}
\plotone{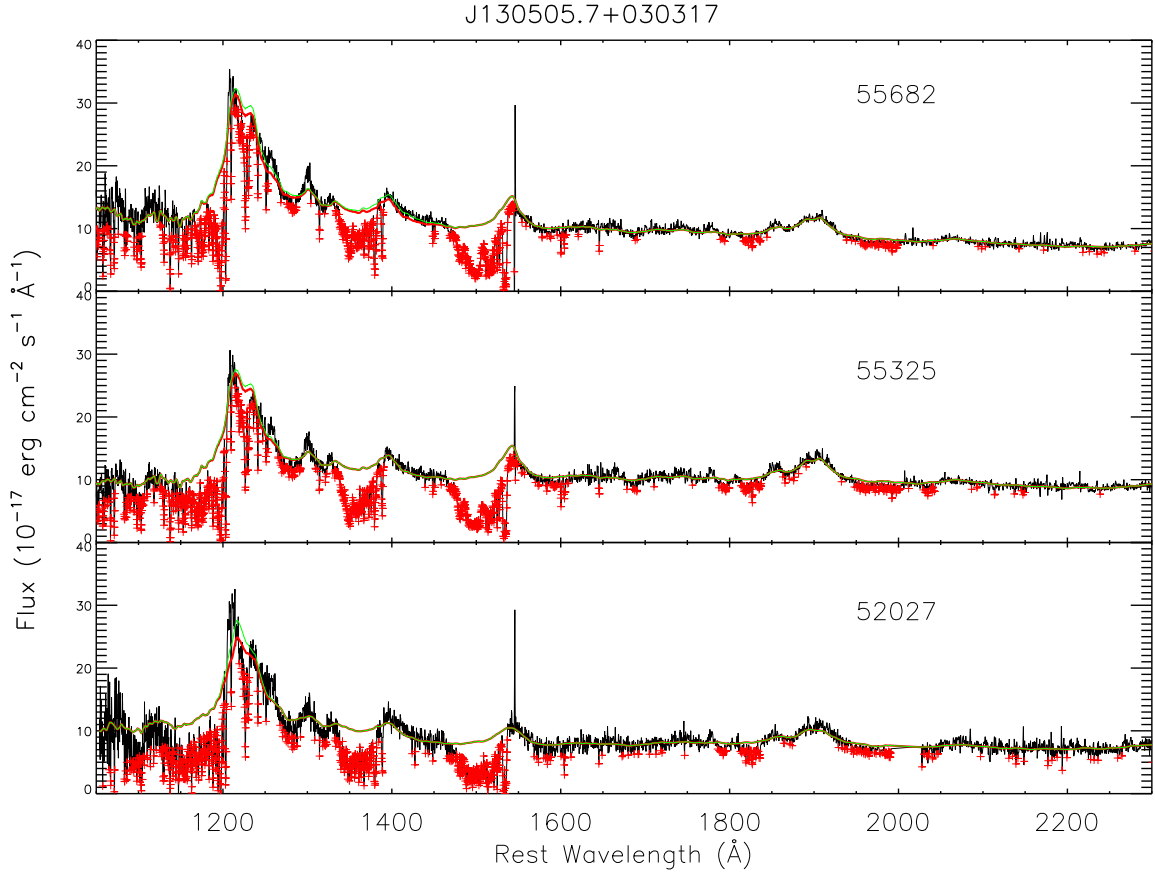}
\caption{Unabsorbed template fit to the spectra of SDSS J130505.7+030317 at three epochs. 
SDSS $mjd$ of the spectrum is labeled in each panel. The red line shows the scaled best 
matched template, while the green line has additional gaussians to account for emission lines. 
The red crosses are those pixels $3\sigma$ below the fit, and thus masked out during 
the fit.
\label{tempfit} }
\end{figure}

We fit these templates to the spectra using a double power-law function as a scale 
factor,
\begin{equation}
S_\lambda=A[1]\left(\frac{\lambda}{2000\AA}\right)^{A[2]}+ 
    A[3]\left(\frac{\lambda}{2000\AA}\right)^{A[4]}
\label{eq1}
\end{equation}
where coefficients $A[1:4]$ were determined by minimizing $\chi^2$. Note that power-law 
can also be used to describe the SMC-like dust extinction in the UV band, which can be 
considered as a good approximation for quasar reddening (e.g., Richards et al. 2003). The 
wavelength range of this fit covers from 1260\AA\ to 2750\AA \footnote{Despite the relative large 
number of templates used, in a significant number of cases, we cannot find a proper 
match for all strong emission lines from 1200\AA~ to 2850\AA.}. We also shift the 
templates in wavelength up to 6 pixels to account for uncertainties in the redshift. 
To do the fit, we iteratively mask  pixels significantly ($3\sigma$) lower than 
the model on the blue side of C~{\sc iv}, Si~{\sc iv} in order to exclude possible absorption 
lines. We select the best ten templates according to the criterion of the maximum 
number of pixels within 1 $\sigma$ error, penalized for the number of pixels exceeding 
the model at $3\sigma$ level. This usually yields a reasonable fit to the SDSS spectrum 
except, in some cases, around strong emission lines. To evaluate the fit around 
strong emission lines, we calculate the significance of excess or deficient 
flux around each line ($\chi=\sum(f-f_{fit})/\sqrt(\sum \sigma_f^2)$), which is 
used as a criterion for the addition of a Gaussian for the excess or deficient 
emission line. In the case of the deficient regions, $\chi$ is only calculated on 
the red side of an emission line for C~{\sc iv}, N~{\sc v} and Si~{\sc iv} because 
redshifted absorption lines are very rare (Hall et al. 2013).  We then extend  it
blueward.  If it is significant at the 3 $\sigma$ level, a Gaussian function is 
then added to the selected best-fit, and a new fit is performed. About 8\% of spectra 
need an additional gaussian component for C~{\sc iv} emission line. In most cases, this 
can reproduce a reasonably good fit as judged by eye. Examples of fits are 
shown in Figure \ref{tempfit}. For convenience, we will term this best fit as ``matching 
template'' although it may include additional gaussians. A normalized spectrum is 
created by dividing the observed spectrum with the matching template. 

\begin{figure}
\plotone{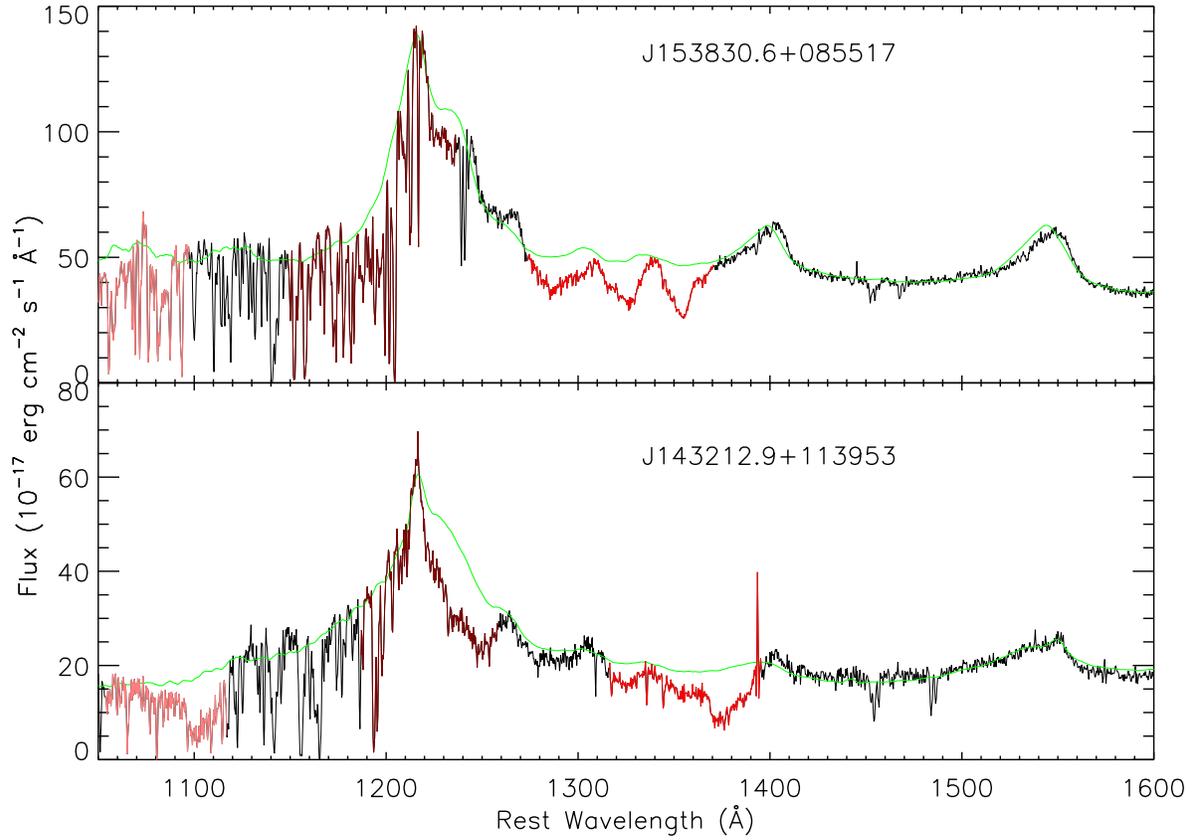}
\caption{Two sampled SDSS spectra of quasars with high velocity C {\sc iv} absorption 
lines at shortward of the Si {\sc iv} emission line. The green line represents the best 
matched unabsorbed template. The high velocity absorption line in C {\sc iv} is shown 
in red; the correspondent portions for Si {\sc iv} and N {\sc v} are displayed in brown 
and in pink, respectively. \label{highv}}
\end{figure}

We searched contiguous deficient pixels for intrinsic absorption lines in 
the normalized spectrum. With a focus on the moderate to broad absorption lines, 
which are more likely intrinsic, deficiency over a width of $\Delta \ln\lambda\geqslant 
10^{-3}$, or 300 km~s$^{-1}$ in velocity, and statistically significant 
than 5$\sigma$, are screened. The significance is defined by the total deficient 
flux divided by the square root of the summed errors.  Then we check by eye to exclude 
false ones, caused by 
an improper fit in most cases. In a number of cases, C~{\sc iv} absorption lines move to 
wavelengths shorter than 1400~\AA, so these may be identified incorrectly as Si~{\sc iv} 
absorption lines. To deal with this, we simultaneously check the absorption lines in 
C~{\sc iv}, Si~{\sc iv} and N~{\sc v} over a velocity range up to 40,000 km~s$^{-1}$ , if 
N~{\sc v} is within the spectral coverage \footnote{In about 10\% cases, 
N~{\sc v} is shifted out of the SDSS spectral coverage because the blueshift of the 
absorption line is too large or because the N~{\sc v} absorption line falls in a 
problematic spectral range as flagged by SDSS mask array by chance.}. Our 
assumption is that any Si~{\sc iv} absorption lines must have an accompanying C~{\sc iv}. 
If not, the feature will be identified tentatively as a high-velocity C~{\sc iv} component. 
As confirmation, we look for a possible corresponding 
N~{\sc v} and Si~{\sc iv} absorption line. Examples of high-velocity absorption lines 
are shown in  Figure \ref{highv}. Note that we do not intend to select a complete 
sample of BAL or mini-BAL quasars, but rather establish a starting point for finding a 
reliable sample of variable absorption line quasars. 

\subsection{Identification of Variable Absorption Lines}

\begin{figure}
\plotone{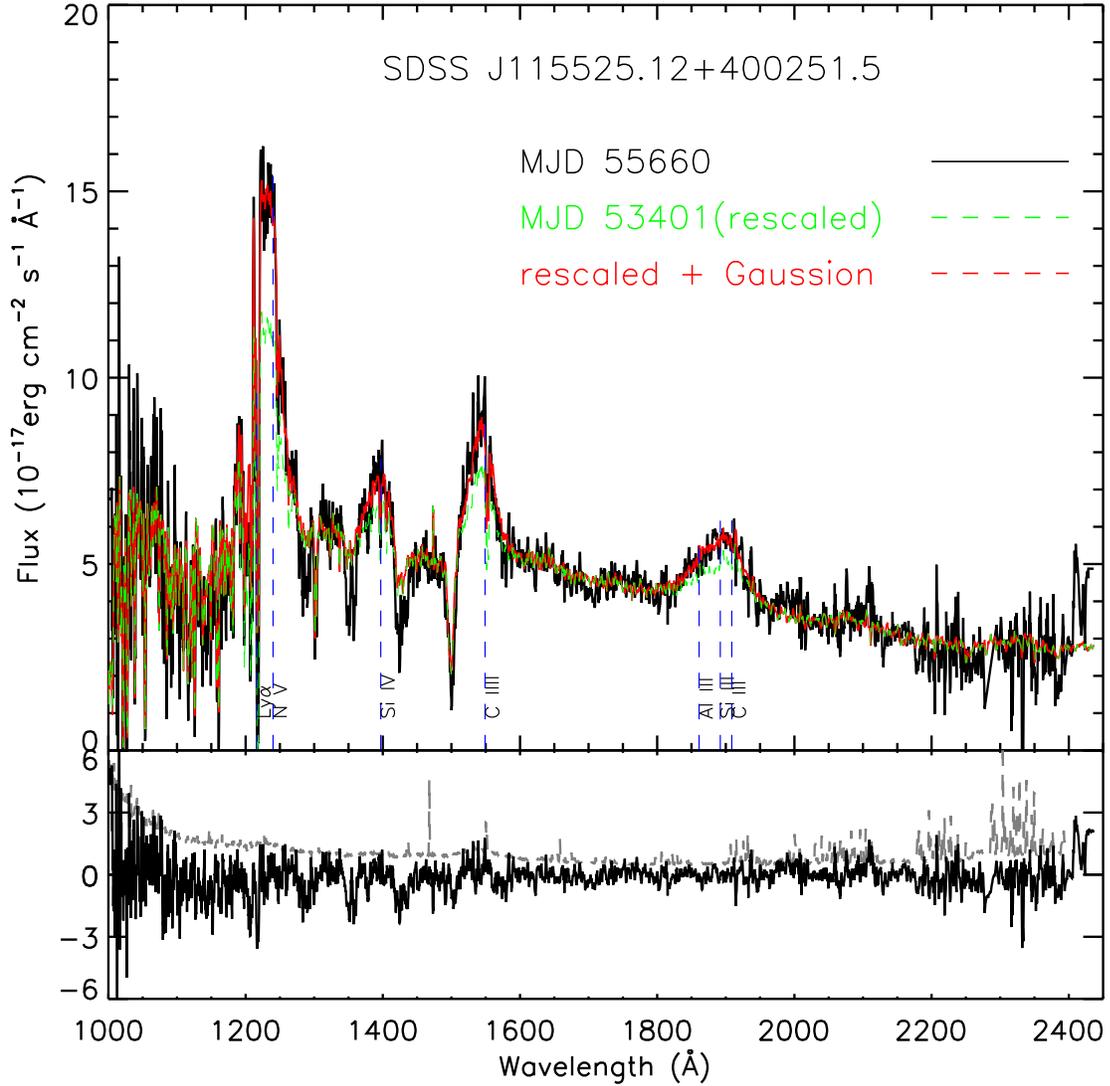}
\caption{Match one SDSS spectrum (in black) to the reference spectrum taking at another 
epoch by multiplying one spectrum with a double power-law described in the text (the green 
line). The red curve represents the one with additional Gaussians to account for the 
change of the emission line equivalent width. The residuals of fits (solid line) and 
the combined spectrum uncertainties (dashed line) are plotted in the lower panel. 
\label{specfit}}
\end{figure}

In order to determine whether emission or absorption lines are variable or not, with 
respect to the continuum, we first select the highest $SNR$ spectrum of the quasar as 
a reference, and then rescale it using the double power-law function 
(Eq.\ref {eq1}), as in the template fit, to match the quasar spectrum in the 
overlapping region obtained at other epochs. This is an empirical approach to take 
into account for potential variations of the continuum shape, in which the spectrum 
usually becomes bluer as continuum brightens, and for the uncertainty in the 
relative spectrophotometric calibration (Dawson et al. 2013 and Appendix A). 
As shown below, with this recipe we can fit the observations very well. Examples of the 
fit are displayed in the Figure \ref{specfit}. 

Once this rescaling is done, we found that weak lines and the continuum SED match very 
well in the two epochs. The median scatter in the difference spectrum is comparable to 
the combined uncertainties (square root of the sum the 
square of the flux errors) of the spectra provided by SDSS in the regions avoiding strong 
emission lines. However, emission lines and absorption lines show significant 
differences in many cases.  
To account for variations of emission line EW, we then add/subtract 
a Gaussian to/from the rescaled spectrum. To minimize potential spurious results, 
the sign of normalization, which will be used as an indicator of the 
increase or decrease of overall emission line EW in section \ref{varpar}, 
is assumed to be the same for all emission lines.  We 
also restrict the center of the Gaussian to lie within $\pm$500 km/s of the line center 
at the source rest frame. Although the actual variation of emission 
lines may be more complicated, for the $SNR$ of our spectra, this 
reproduces good fits for almost all lines. With this fit, we found that the 
scatter in the difference spectrum over emission-line regions is similar to that 
of continuum regions. In comparison with unabsorbed quasar template 
matching, the rescaled reference matching usually produces a better fit outside 
the absorption line region. So in the following, we will measure the absorption 
line variability from the difference spectrum rather than by comparing EWs 
obtained in the template fitting.

To identify variable absorption-line components in the difference spectrum, we 
searched for contiguous negative and positive bins to determine the range of 
variable components. The uncertainties are estimated by taking into 
account the flux uncertainties of the two spectra given by the SDSS pipeline, and 
possible systematic uncertainties due to rescaling. The latter is estimated in 
neighboring regions that are outside of the absorption troughs.
Because difference spectra do not have a high $SNR$, 
in general,  a single variable component may split into two or more segments 
due to statistical fluctuations in the spectrum. 
To overcome this problem, we take three steps. First, we mark all pixels where the 
difference is larger than 5\% of the average value and more than 3$\sigma$. Adjacent marked pixels 
are then connected to form a variable region. Next we expand such regime into neighboring 
pixels that have the same sign in the difference spectrum but at the less than 3 $\sigma$ significant 
level. After that, we also merge the neighboring regions with the same variable sign 
and with a separation of less than four pixels. Finally, we identify the same 
component in different epochs, and then extend the velocity range of each component 
to cover the component at all epochs. 

\begin{figure}
\plotone{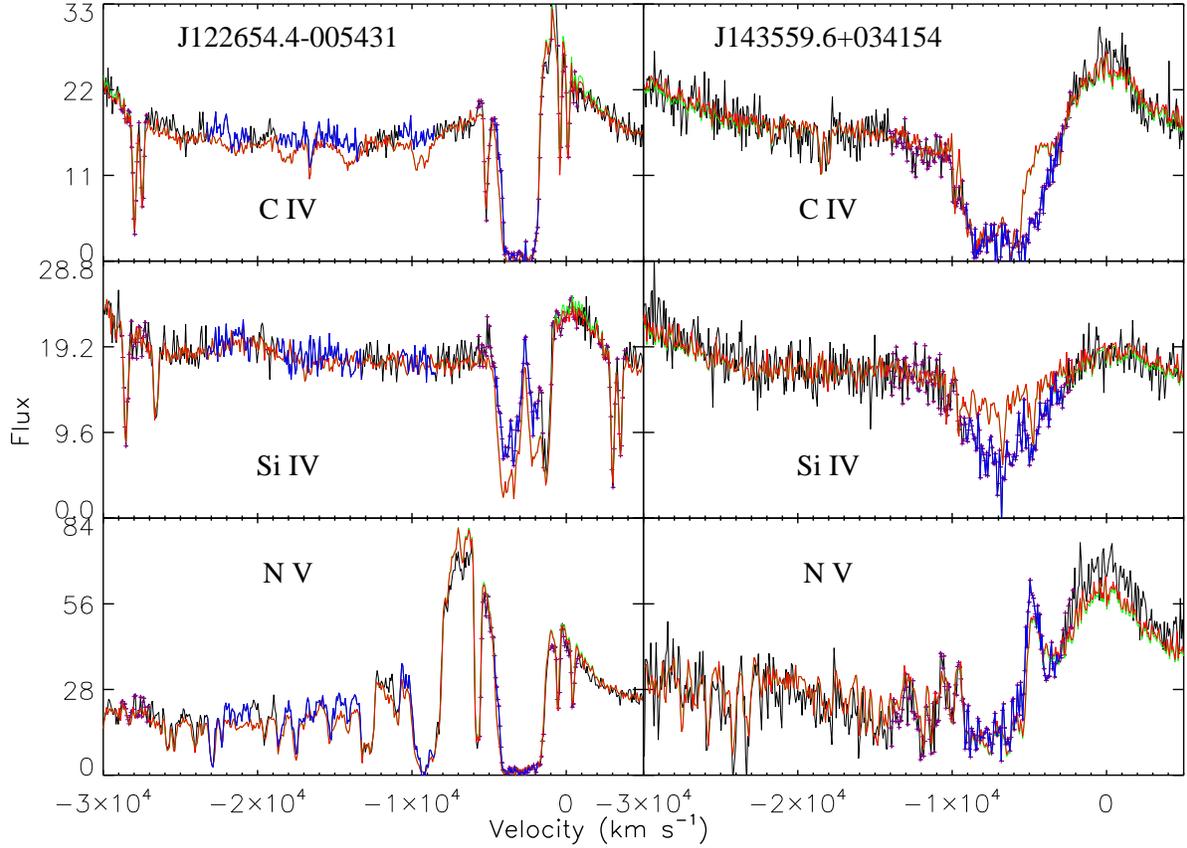}
\caption{Different variations of C {\sc iv}, Si {\sc iv}, and N {\sc v} absorption lines 
in two quasars. The variable region in one or more absorption lines are marked in blue.
Other colors are the same as Figure \ref{specfit}. Noting strong C {\sc iv} and N {\sc v} 
absorption lines are not apparently variable in the deep trough while the weaker absorption 
line Si {\sc iv} does. This is likely an observation effect that it is more difficult to 
detect small flux changes in the deep absorption trough caused by ionic column density 
variations. The absorption trough around -9000 km~s$^{-1}$ in lower-left panel is due 
to Ly$\alpha$ absorption. \label{largedepth}}
\end{figure}

In practice, we iterate three times through the above procedures. We mask the 
variable absorption regions identified in the previous iteration, rescale the spectrum, 
and refit the emission lines. After that, we redefine the variable absorption-line 
region. Finally, the sample is then examined by eye and spurious regions are excluded. 
In the end, a sample of 452 quasars with detected variable absorption lines were identified.
We list the sources and regimes of the variable absorption lines in table \ref{tab1}. 
The maximum velocities of the C {\sc iv} absorption line troughs in  this 
sample extend to up to 45000 km~s$^{-1}$, and the widths of individual absorption troughs 
are in the range of 500 km~s$^{-1}$ to 15000 km~s$^{-1}$. Note that variable components may 
be the entire trough or just a portion of the trough. The 
latter does not necessarily mean that the variable portion of the trough is physically 
different from the rest, instead it may reflect the limitation that we cannot detect 
small variations caused by even moderate changes in the ion column density in certain 
circumstances. This is evident especially in the deep trough of  C {\sc iv}, where 
we sometimes observe an apparent variation of the Si {\sc iv} absorption line while no 
similar C {\sc iv} variation is seen (Figure \ref{largedepth} and also Filiz Ak et al. 2013). 
In comparison with those of Filiz Ak et al. (2013), our sample also includes absorption lines 
that do not meet the criteria of BALs. The basic properties of the sample are summarized 
in the Table \ref{sample}.

\section{Measuring the Variability Parameters and Statistical Method \label{varpar}}

In this paper, we will use the sign of the variation with respect to the scaled reference spectrum 
as a non-parametric description of absorption line variability. We simply measure the integrated 
variable flux over the variable region of absorption lines in the difference spectrum using the 
reference matching method. We assign a sign of absorption line variation of $+1$ when the 
absorption line trough becomes significantly deeper (at the 3 $\sigma$ level),  $-1$ when the trough 
gets significantly shallower, and zero otherwise. In comparison with previous measurements
using  variations of an absorption line EW, our method avoids the additional uncertainty 
introduced in the process of template matching while measuring the EWs of absorption 
lines in each spectrum.  

We estimate the variability of the continuum and emission lines in order to 
explore the internal driver for absorption-line variability. It is known that 
the spectrophotometric calibration of SDSS has an uncertainty of about 
$\sigma_r=0.05$ magnitudes in the SDSS I/II surveys (Abazajian, et al. 2009) and 
is considerably worse for BOSS spectra (Dawson et al. 2013; Par\'is et al. 2014). 
However, as shown in Appendix A, the distribution of difference magnitudes between 
two epochs of our detected variable absorption line quasars is considerably broader 
than possible calibration uncertainties.  Hence it is still possible to extract statistically 
useful information about the sign of continuum variability. We consider a continuum 
variation significant when its amplitude is greater than 5\% at the $g$-band, or 
typically, $\sim$1400\AA~ in the quasar rest frame. As for absorption lines, we assign 
a sign of the continuum variation to $+1$ when continuum flux around 1400\AA~ is more 
than 5\% brighter than the reference spectrum, $-1$ for the opposite case, and $0$ 
otherwise.

If there are only two categories, the concordance index should follow 
a binomial distribution with a probability of $P(x>k;n,p)=\Sigma_{i\leq k} C_n^i 
p^i(1-p)^(n-i)$, for a result of more than $k$ concordant cases in $n$ pairs 
with a concordant probability of $p$, and anticoncodant probability of $1-p$ 
for each pair, where $C_n^i$ is the binomial coefficient. In our case, the 
concordance index can take an additional value $0$ because of an undefined 
variability sign of one variable, which does not affect our analysis. 
Because undefined variability signs can be caused by a low signal to noise 
ratio of spectrum or small variability, which are not interested to us, we 
will ask the probability $P(x>k;n,p)$ among $n$ those pairs with defined 
variablity signs of both variables. This probablity follows a binomial 
distribution. If we take $p=0.5$, then $P$ gives the chance coincidence 
that concordance and anti-concordance have equal possibility. Furthermore, 
one can derive the most likelihood value 
of $p=k/n$, and 1 $\sigma$ error in $p$ as $\sqrt{p(p-1)/n}$ if there is no 
measurement error in the concordance index. On the other hand if the concordance 
index were mis-assigned randomly for a fraction $q$ of sources, we can 
derive easily $p=(k/n-q)/(1-2q)$ by considering those mis-assigned events. 
The error can be estimated by Monte-Carlo simulations. It is evident that taking 
into account of mis-assigments will make $p$ larger if $p>0.5$ and smaller if 
$p<0.5$. This is reasonable because random errors would only smear the difference 
rather than to enhance it, making apparent $p$ closer to 0.5.

In view of the well-established intrinsic Baldwin effect, i.e., the EW of an 
emission line correlates negatively with that of continuum during 
continuum variations (e.g., Kinney et al. 1990), we will use the variation 
of emission-line EW as an independent check for the sign of continuum variation. 
Lags in the emission line to continuum variations  will introduce a correlation 
similar to the Baldwin effect. We calculate variations of the EW of strong 
emission lines (including Ly$\alpha$, N~{\sc v}, Si~{\sc iv}, C~{\sc iv}, C~III]). 
The variation of emission lines relative to the continuum is measured in the 
reference-matched spectrum, and the sign of emission line variations can be directly 
extracted from the sign of additional Gaussian components in the reference matched 
spectrum. Similarly, we assign a sign for the emission line variation: +1 if line 
flux in the reference matching spectrum increased significantly; and -1 if line 
flux decreased significantly; and 0 otherwise. 

Among 510 spectrum pairs of 452 objects (Table 1), 292 pairs show opposite signs of 
continuum and emission line $EW$ variability; 111 pairs display the same sign; and rest 
107 pairs does not have well defined variability sign of either continuum or emission 
lines. So in a total 403 cases with well defined variability sign in both continuum and 
emission lines, 292 has concordance index $-1$. Following the binomial distribution, 
we obtain a probability of concordance index $-1$: $p=0.725\pm 0.022$, which is 
10.2$\sigma$ from no preference ($p=0.5$), for the anticorrelation between continuum 
and emission line variability. So the intrinsic Baldwin effect is detected at a high 
conficence level for the high luminousity quasars.

\section{Correlated Variations of Absorption Lines, Emission Lines and Continuum} 

\subsection{Correlated Variations of Absorption Lines} 

Previous observations have shown that different components of the C~{\sc iv} absorption 
line show coordinated variations, i.e., in the same sign of variations (see 
references in \S 1). This is confirmed by our study. We define a concordance 
index with $+1$ for the all components that vary with the same sign and $-1$ for one 
or more components with opposite signs. Among 114 spectral pairs of 101 objects 
with multiple variable C~{\sc iv} components, only 25 pairs display one or more 
components varying in opposite sign to others, while the rest show the same sign. 
This gives a probability for concordant variations of $p=0.781\pm 0.039$, which is 
7.2 $\sigma$ from $p=0.5$ 
following the binomial distribution. Note that, when only the most significant variation is 
selected for each object, if there are more than one spectral pairs, 97 of 101 
objects show concordant variations of different components, consistent with the 
above conclusion. Furthermore, we notice that in the case of incongruent variations, 
the two asynchronous components often share an absorption trough, indicating an 
acceleration or deceleration of an absorption-line component. This will be studied in 
detail in a forthcoming paper.  

\begin{figure}
\plotone{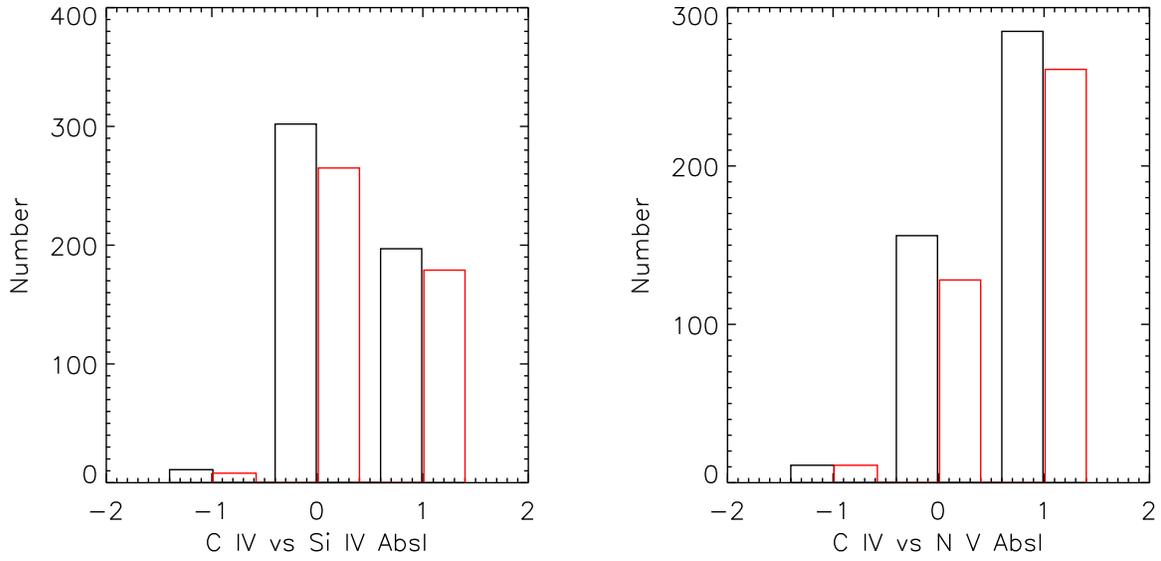}
\caption{The distribution of concordance index for the same variable component of 
different lines (left panel: C~{\sc iv} versus Si~{\sc iv}, right panel: C~{\sc iv} 
versus N~{\sc v}).  The concordance index is defined as +1 if both lines vary in the 
same sign (increasing or decreasing), -1 in opposite sign and 0 otherwise. We exclude 
58 spectra pairs and 52 quasars from analysis that N~{\sc v} absorption lines are outside 
of spectral coverage. The black boxs are 
the distribution of all spectra pairs in table 1, while the red boxs are based on 
the most significant pair for each object. 
\label{c4si4}}
\end{figure}

Next, we explore the coherence of the variations of the same component of different 
absorption lines. It was demonstrated by several previous authors that variations 
of C~{\sc iv} and Si~{\sc iv} absorption lines are highly concordant (see references 
in \S 1). Figure \ref{c4si4} shows the distributions of the concordance index for 
(C~{\sc iv}, Si~{\sc iv}) and (C~{\sc iv}, N~{\sc v}) pairs.  In cases of 
more than one variable C~{\sc iv} absorption line component, for each line we 
summed up the sign of individual components, and assign a sign $+1$ for a positive 
sum, $-1$ for a negative sum and 0, otherwise. Using this sign, we calculate the 
concordance index for the spectrum. We do count different spectra of one quasar 
as independent. However, absorption lines that are affected by bad pixels or are out 
of the spectral coverage were not counted. From these distributions, 
there are only a small fraction ($<$4\%) of cases where the two lines vary in an opposite 
way although there is still a large fraction where Si~{\sc iv} or N~{\sc v} does not 
vary significantly. C~{\sc iv} and N~{\sc v} are correlated better than C~{\sc iv} and 
Si~{\sc iv}, i.e., 28\% of pairs have concordance index of zero for C~{\sc iv} vs N~{\sc v} 
while 60\% between C~{\sc iv} and Si~{\sc iv}. Using the distributions in Figure \ref{c4si4}, 
we calculate $p=0.944\pm0.015$ (black boxes; $0.949\pm0.017$ for red boxes) for concordant 
variations of C~{\sc iv} and Si~{\sc iv} absorption lines. The probablity for concordant 
variation of C~{\sc iv} and N~{\sc v} is $p=0.963\pm0.010$ for the black boxes in the 
right panel of Figure \ref{c4si4}, and $0.960\pm0.012$ for the red boxes.
 
A close examination of the cases of concordance index zero shows that, in most cases, the 
non-variable line (N~{\sc v} or Si~{\sc iv}) either has a very shallow or very deep absorption 
trough. So the non-detection of variability is most likely due to the insensitivity of the 
variation of the flux in absorption trough to changes in the ion column density in these 
cases, and the relatively low $SNR$ ratio of SDSS spectrum. This explains why more sources 
in the Si~{\sc iv} and C~{\sc iv} pairs have an concordance index of zero than C~{\sc iv} 
and N~{\sc v} pairs. The latter pairs have similar optical depths while Si~{\sc iv} usually 
has a considerably smaller optical depth. In fact, Si~{\sc iv} variations are observed more 
frequency in  quasars with a deep C~{\sc iv} absorption trough. We will explore the 
implication of this in the next section. 

\subsection{Coordinated Variations between Absorption Lines and Continuum/Emission Lines}

\begin{figure}
\plotone{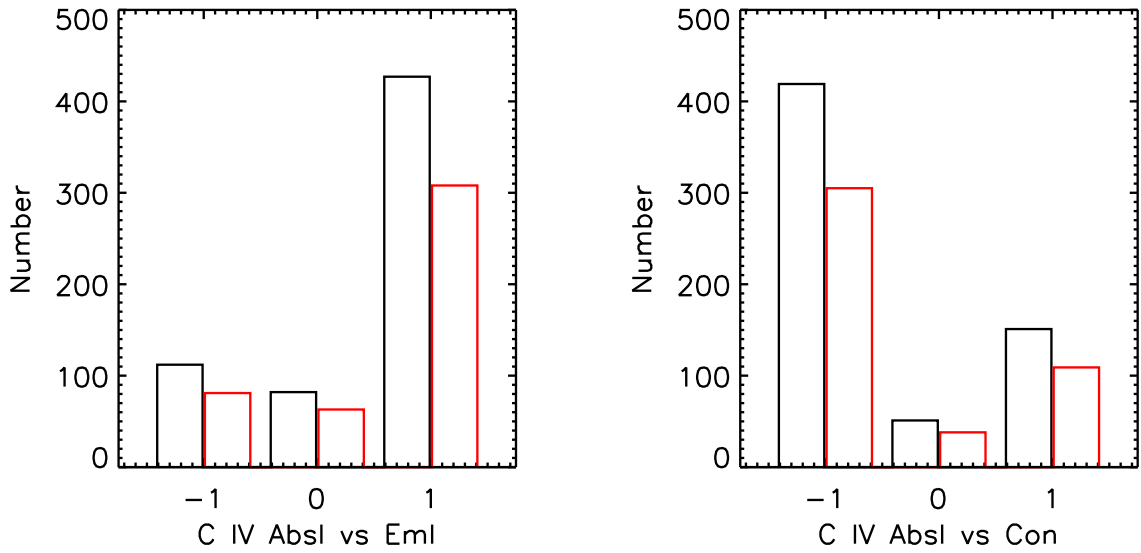}
\caption{The distribution of concordance index for C~{\sc iv} absorption line versus 
continuum (right panel), and  C~{\sc iv} absorption line versus emission line (left panel). 
The concordance index is defined as +1 if the absorption and emission lines weaken or 
strengthen simultaneously, -1 if one weakens and the other strengthens, and 0 otherwise. 
Similarly, in the case of continuum versus absorption line, a concordance index +1 means 
absorption line strengthens while continuum brightens and vice versa. Only quasars with 
variable C~{\sc iv} absorption lines are included. See text for 
definition of this index. Colors are as in Figure \ref{c4si4}.
\label{concordance1}}
\end{figure}

Correlations between variations of absorption lines and continuum/emission lines 
provide important insight into the origin of absorption-line variability. However, 
previous studies have not reached a concensus on whether absorption line and continuum 
variations are correlated or not. Barlow (1994) showed marginal evidence for a 
correlation, while Gibson et al. (2008) found that the correlations between the variations 
of absorption line equivalent width of C {\sc iv} and continuum flux is not significant 
for 14 BAL quasars. In a recent study, Vivek et al. (2014) did not find any trend in the 
variation of the Mg~{\sc ii} and Al~{\sc iii} BAL equivalent widths with the variations 
of continuum parameters. In a recent study, Wildry et al. (2014) argued that variations 
of Si~{\sc iv} and C~{\sc iv} BALs are not driven by photoionization based on the lack 
of correlations with continuum luminosity. We only consider the signs of absorption, 
emission line, and continuum variations in this subsection, because the exact amplitude 
of continuum variability is rather difficult to quantify with only SDSS spectra in 
view of uncertainties in the absolute flux calibration. Since C~{\sc iv}, Si~{\sc iv} 
and N~{\sc v} absorption lines show the same sign of variations, for each absorption c
omponent, we count only the sign of significant variation in 
the C~{\sc iv} line, and compare this with continuum variations. As in last subsection, 
we assign a concordance index $+1$ for the case where an absorption line becomes stronger 
when continuum brightens, and  $-1$ in the opposite way, while $0$ for the case in 
which the sign of continuum variation is not determined. We also define a 
similar concordance index to quantify the coherence between the variations of 
absorption lines and emission line EWs. We show our results in Figure 
\ref{concordance1}.  

Obviously, the variations of absorption lines are statistically highly coordinated 
with those of continuum, and EWs of emission lines. The apparent probability for 
concordant variations of emission and absorption line is $p=0.796\pm 0.017$, which is 
17.4$\sigma$ for $p=0.5$. We get the same result if only the most significant variation 
for each source is considered (using the red histogram of Figure \ref{concordance1}. 
According to the intrinsic Baldwin effect, the two results suggest that absorption 
lines weaken as the continuum brightens, and vise-versa.
Similarly, the apparent probability for concordant variations of continuum and 
absorption line is $p=0.726\pm0.016$, which is 14.1$\sigma$ from $p=0.5$. We obtain 
a similar result $p=0.726\pm0.022$ (10.3$\sigma$) when only the most significant 
variations are considered for each source.  Considering possible mis-assigned signs 
for the variations of either variables, the real $p$ should be higher. 

The conclusion on the correlated variations of absorption line and continuum should 
be not affected by the SDSS flux calibration uncertainty of SDSS spectra.
{\em Because the spectrophotometric 
calibration error is not expected to correlate with the variation of the absorption-line 
strength, any calibration errors will only diminish rather than enhance the correlated 
variations. Thus, the observed coordinated variations of absorption lines and continuum 
must be real. This is further supported by coordinated variations between absorption 
line and emission line EW, which are not 
affected by the flux calibration.} Note that we use the continuum flux variability 
amplitude of $>5\%$ as a threshold, while it is estimated conservatively in appendix 
A that about 13\% have been assigned to an opposite sign of continuum variations, and 
additional 10\% with undefined sign of variations.    
The total number is comparable to the total fraction (33\%) of sources with a concordance 
index of 0 or 1. Thus, our result is consistent with the idea that a large fraction 
of cases with a concordance index of 0 or $-1$ are due to absolute flux calibration 
uncertainties. If we take the numbers given above, about 14.4\% of BAL quasars 
were assigned to a wrong sign of continuum variation among these with defined signs. 
According to equation in \S \ref{varpar}, we obtain $p=0.817$. However, the exact 
probability has to wait till a complete solution of the flux calibration uncertainties 
are found. We see that there 
are fewer cases of concordance index $-1$ in the absorption and emission line pairs because 
there is no such calibration problem, although the measurement of the variation of 
emission-line EW is subject to larger statistical uncertainties.

\begin{figure}
\plotone{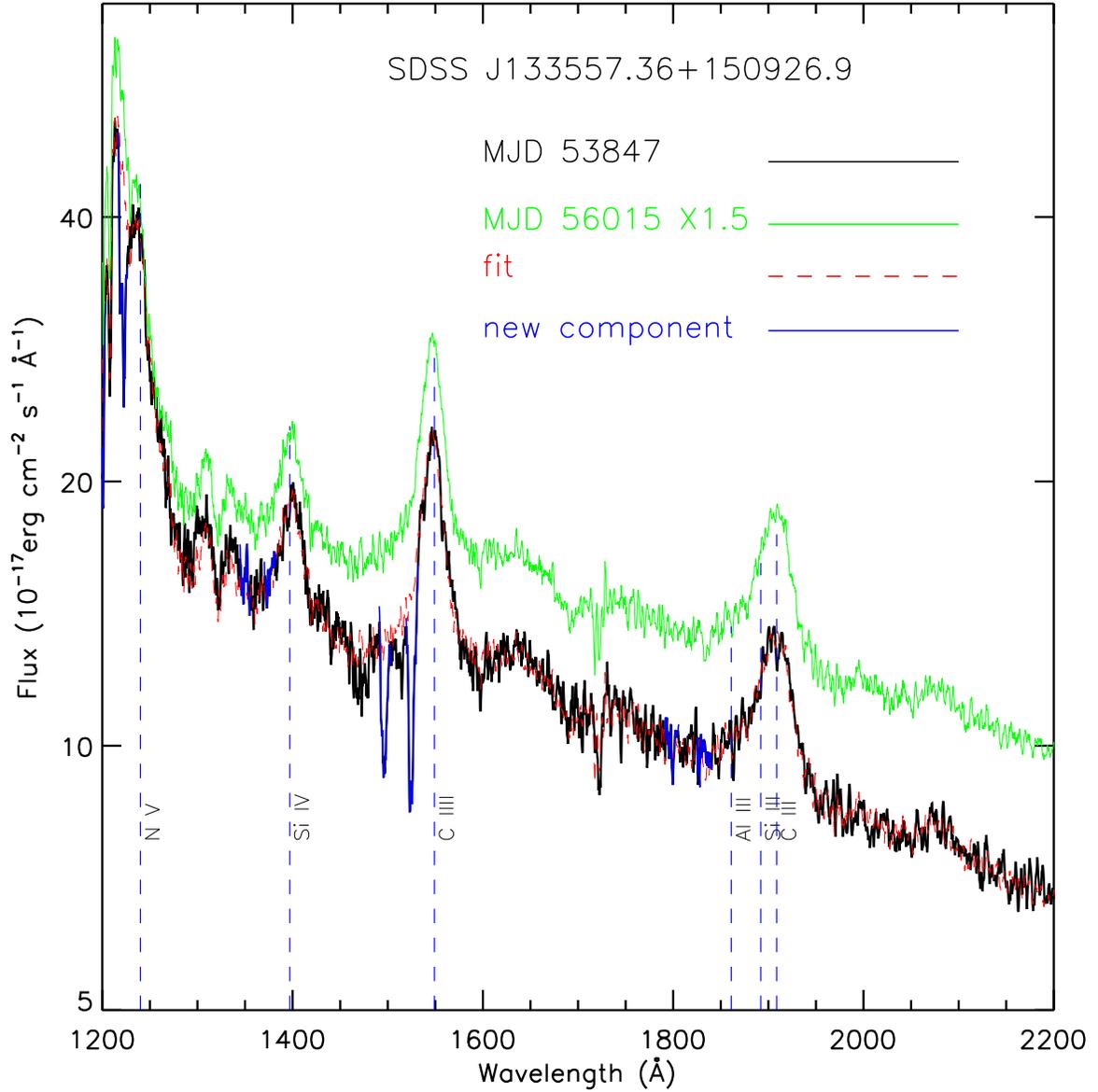}
\caption{Example of emergence or disappearance of new BAL components. The SDSS 
spectra at two epochs are shown as black and green lines, while the best fitted 
template is overplotted in red. Two new C~{\sc iv} absorption line components are 
marked in blue. The corresponding spectral region of Si~{\sc iv}, N~{\sc v} and Al~{\sc iii} 
are also shown in blue. The new absorption line components are also visible in N~{sc v}.
\label{newbal} }
\end{figure}

\begin{figure}
\plotone{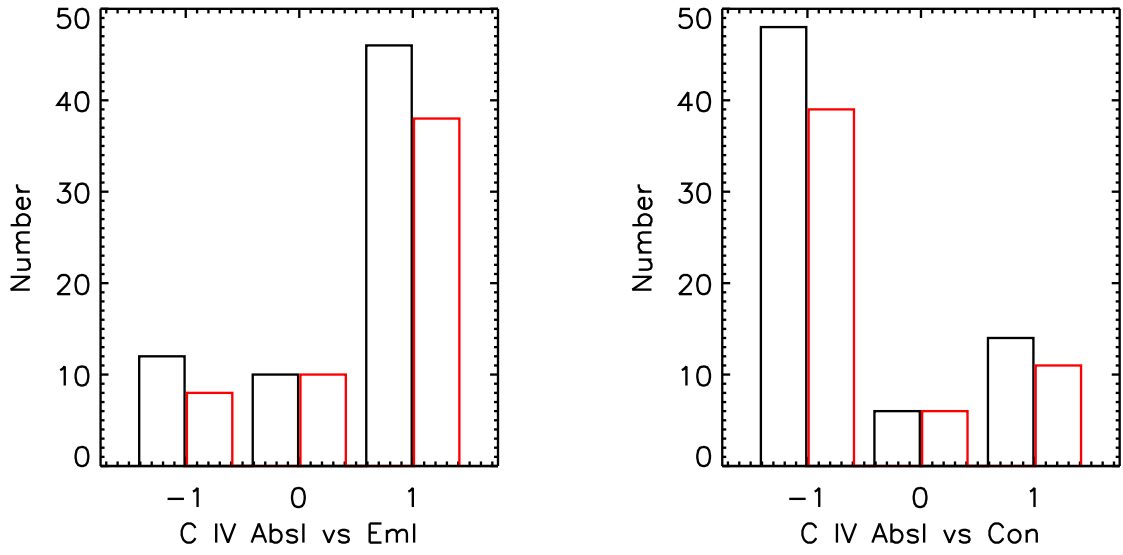}
\caption{The distribution of concordance index between the C~{\sc iv} absorption line and 
continuum (right panel) and  C~{\sc iv} absorption line versus emission line (left panel) 
for quasars with one or more new BAL component. The black and red boxs are the same 
as in Figure \ref{c4si4}. \label{concordance2}}
\end{figure}

There are rare cases where a new component of an absorption line emerges or a 
previously known component disappears (Hamann et al. 2008; Leighly et al. 2009; Krongold, 
Binette, \& Hern\'andez-Ibarra, 2010; Filiz Ak et al. 2012). They are extreme cases 
where the absorption-line trough is below the detection limit in one epoch, which 
depends on the $SNR$ of the spectrum and the correctness of the template 
used to match the spectrum. These cases are often taken as evidence for outflows 
moving in and out of the line of sight. The large sample of multi-epoch spectroscopic 
quasars enables us to extract a statistically-significant sample of about 
70 spectral pairs of around 50 quasars with emergence or disappearance of at least one 
absorption line component (see Table \ref{tab1} and Figure \ref{newbal}). We performed 
the same analysis of the concordance index among variations of continuum and absorption 
lines, or emission and absorption lines. The results are shown in Figure \ref{concordance2}. 
The coordinated variations of absorption lines, continuum, and emission lines are 
statistically significant at 5.5 (for all pairs, black boxes in the Figure) 
or 5.8$\sigma$ (only most significant pair for each object, red boxes) and 5.2 (all pairs) 
or 4.8 $\sigma$ (only one pair for each object), respectively. 
It is apparent that these cases also follow the above correlations, suggesting that 
the same physical process also accounts for these extreme variations of ion column 
density. Because they account for only a small fraction of our variable absorption 
line quasar sample, they do not affect the statistical results of the parent sample 
presented above.

\section{Discussion}

\subsection{A Short Summary of Previous BAL Variability Studies and Our Results}

Previous works found that variations of BALs on the time scales of years are 
common (50-60\%), and that the variable fraction and variability amplitude increases with 
the time interval of observations (Gibson et al. 2008; Capellupo et al. 2012; 
Filiz Ak et al. 2013). These authors also found that variations usually take place only 
in a narrow and relatively shallow portion of a broad BAL trough although it is 
not clear how much this is due to saturation of the absorption line (see also 
Lundgren et al. 2007). 
The distribution of EW variations can be decribed using a random walk model (Filiz Ak 
et al. 2013). It is also established that variations 
of C {\sc iv} and Si {\sc iv} are highly coordinated so are variations of different 
BAL components of C {\sc iv} (Capellupo et al. 2012; Filiz Ak et al. 2012, 2013). 
Disappearance or emergence of BAL components are observed in a small fraction of 
BAL quasars on time scale of years (Hamann et al. 2008; Leighly et al. 2009; 
Filiz Ak et al. 2012; Rodriguez Hidalgo et al. 2013), and Filiz Ak et al. 
(2013) suggested that these are the extreme case of BAL variability rather than a 
new phenomenon based on the overall distribution of variability fraction of 
the absorption line equivalent width. 

In this paper, first, we extended the coordinated variations of different 
lines to N {\sc v} and C {\sc iv} absorption line. Second, we showed that 
variations of absorption lines are highly coordinated with continuum 
variations. Finally, the emergence or disappearance of an absorption line 
component is correlated with dimming or brightening of UV continuum. 
Regarding the second point, it should be poined out that data in previous 
studies are consistent with our results although they were interpreted in a 
different way some times. Barlow et al. (1992) found a marginal evidence 
for a correlated continuum and absorption line variations, 
Gibson et al. (2008) did not find significant correlations 
between the variations of the EW of CIV BALs with those of continuum for a 
sample of 13 BAL quasars observed in LBQS and SDSS. However, their result does 
not contradict ours. Among 13 objects, 7 objects lie on the first and third 
quadrants, and 3 on second and fourth quadrants on $\Delta EW$ vs $\Delta 
\log F(2500\AA)$ diagram (their Figure 10), and the other three do not have 
significant variations in either one parameter or both (less than their errorbars). 
The situation is quite similar on the $\Delta EW$ vs $\Delta \log F(1400\AA)$ 
diagram, among 7 objects with significant variations of both parameters, 5 fall 
on the first and third quadrants and 2 on the second and fourth quadrants. The 
insignificance is simply due to the small number in the sample. Although Vivek 
et al. (2014) did not found any significant trend between the variations of 
absorption lines (Mg~{\sc ii} and Al~{iii}) and continuum flux, in their Table 
3, 6 of 8 quasars showed opposite trend of long term (on scales of years) 
continuum and absorption lines, only one displayed the same trend and another 
one exhabited both weakening and strengthening BAL components. In this section, 
we will discuss the implication of these findings.

\subsection{What Drive Variations of Absorption Lines?}

Absorption line variability can be caused by changes in either the gas ionization or 
in the total gas column density. This can be produced by two main processes: changes in 
the ionizing continuum incident on the absorbing gas, or gas moving in or out of the 
line of sight (Smith \& Penston 1988; Barlow et al. 1989; Barlow 1993). Previous 
studies did not reach a consensus on which one is the main driver. On the one hand, 
coordinated variations of different components of absorption lines suggest that the 
changes in different part of the outflows are driven by the same parameter. Most 
naturally, this parameter is the ionizing continuum because it ionizes all gas 
producing these absorption components (Hamann et al. 2011; Capellupo et al. 
2012). It is difficult to image that line-absorbing gas at very different sites 
moves in or out of the line of sight in step unless the flows are confined to a 
relatively small region. The latter 
is less likely, as components with large velocity differences will spread out on a 
relatively short time scale, but can not be rejected completely. On the other hand, 
there are cases where a new C~{\sc iv} absorption line component emerged or an old 
component disappeared. Such extreme variations can be readily explained as absorbing 
gas moving into or out of our line of sight (Hamann et al. 2008; Leighly et al. 2009; 
Krongold, Binette, \& Hernández-Ibarra, 2010; Hall et al. 2011; Filiz Ak et al. 2013). 
Capellupo, Hamann \& Barlow (2014) argued that the similar variability trends 
of P {\sc v}, C {\sc iv} and Si {\sc iv} in Q 1413+1143 disfavoured the continnuum 
variations as the driver of BAL variability in this quasar.

Changes in the ionizing continuum striking the absorbing gas may be caused by the 
variations of the intrinsic ionizing continuum or/and by alteration of the column 
density or ionization of an intermediate absorber (refer to discussion in 
Misawa et al. 2007; Filiz Ak et al. 2012, 2013), 
i.e., the shielding gas introduced to by Murray et al. (1995). According to their 
model, shielding gas is optically thin to the observed UV continuum longward of 1000\AA. 
Thus changes in 
the column density of shielding gas would not directly induce a correlated variation 
between UV continuum and absorption lines. On the other hand, if the main change in 
the ionization of shielding gas is caused by an increase or decrease of the 
ionization state in response to variations in the ionizing continuum, it will 
introduce a correlation between absorption lines and observed UV continuum. But 
in this case, the ultimate driver of the absorption line variability is the 
variation of the ionizing continuum.

The correlated variations between absorption lines and the continuum found in 
this paper directly support the idea that the intrinsic ionizing continuum is 
the main driver for absorption-line variability (see Barlow et al. 1992). We also 
showed that emergence 
or disappearance of an absorption line component is closely correlated with the 
continuum variations, suggesting that they are caused by ionizing continuum 
variability as well.  {\em This indicates that there was/is persistently 
absorbing gas}, but the ionization change is so large that the absorption-line 
emerges/disappears in the later epoch. We point out that large X-ray absorption 
column density accompanying with the transient broad absorption lines (Kaastra 
et al. 2014) cannot be simply explained by an ionization change, but rather are 
more likely due to the launching of a new outflow component or inserting of a new 
component of shielding gas. So, gas moving in and out may happen sometimes, but 
it is not the dominant mechanism driving quasar absorption line variability.

\subsection{Implication for Concordant Continuum and Absorption Line Variations\label{implication}}

While highly concordant variations of absorption lines and the continuum strongly 
suggest that the response of gas ionization is to the ionizing continuum, it 
should be noted that this concordance is not an inevitable result of the latter.
The ionic column density of a specific species may respond to a 
continuum variation, positively or negatively, depending on the ionization of 
absorbing gas. 
This is illustrated in Fig \ref{models}, where the fraction of $C^{3+}$ 
increases first, then reach a peak, and decreases after as ionization parameter 
increases. Moreover, if the outflow spans a large ionization range, one would 
expect that different parts may response to the continuum variation differently.
The coherence of different components indicates that the ionization 
of the outflows is globally higher or lower than the critical value that separates 
positive and negative response. The fact that the depth of absorption lines is 
inversely correlated with the continuum flux implies that species more highly 
ionized than that producing the observed absorption lines always dominate. A 
further constraint can  be drawn from the coherent variations of different 
absorption lines, where Si~{\sc iv}, C~{\sc iv} and N~{\sc v} show the same 
variation sign. For instance, the gas ionization must be so high that most 
nitrogen is in ionization stages
higher than N$^{4+}$. Similarly, the statistical concordance of the 
variations of absorption lines and the continuum suggests that in most objects, 
outflows are highly ionized. To put further constraints on the ionization 
of the absorber, it would be interesting to check whether O~i{\sc vi} or 
Ne~{\sc viii} lines also vary in synchrony with other lines, which is beyond 
the scope of this paper. If this is indeed the case, these UV absorbers 
may be also responsible for the warm absorption observed in X-ray band. 
The connection between X-ray and UV absorbers has been discussed for a 
long time, and it was proposed that at least some of the X-ray and UV absorbing 
material is physically connected in some Seyfert galaxies based on either 
similar absorption line profile or correlated temporal variations (e.g., Kaspi 
et al. 2002; Gabel et al 2005; Kaastra et al. 2014) although the exact relation 
is not clear.   

Alternatively, the outflow may be multi-phase with a range of ionization states, as 
recently proposed by Arav et al. (2013) based on analysis of far UV absorption lines 
of the quasar HE 0238-1904, or as known in warm absorbers of Seyfert galaxies (e.g, 
Steenbrugge et al. 2009; Detmers et al. 2011). If the higher ionization phase dominates, 
then the competition between the positive response of ionic column density to the 
continuum variations in low-ionization material and the negative response in high 
ionization gas may finally lead to the observed negative response in the BAL 
quasars. Detailed photoionization simulations are required to test whether this is a 
physically possible scenario and to constrain the physical parameter range. 

\begin{figure}
\plotone{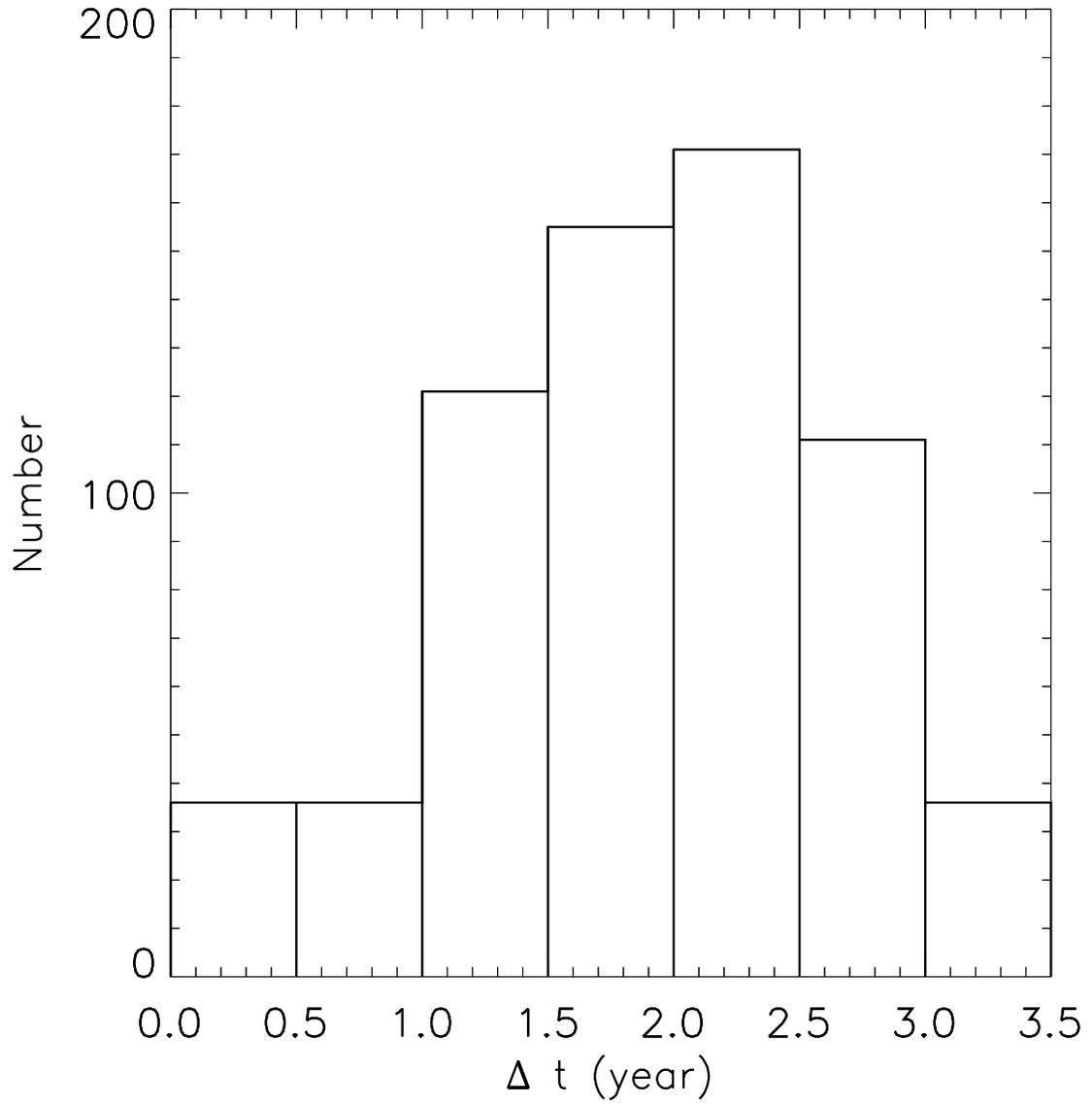}
\caption{The distribution of time interval in the quasar rest frame between the two 
observations with the detection of absorption line variations. \label{interval}}
\end{figure}

The high concordance between continuum and absorption line variations also requires a 
recombination time $t_{rec}=(n_e \alpha)^{-1}$ shorter than both the time scale of 
typical continuum variations and the interval between the two observations 
(e.g., Barlow et al. 1992). At a nominal temperature of 20,000K (Hamann et al. 1997), 
the recombination rates for C~{\sc iv} to C~{\sc iii} and N~{\sc v} to N~{\sc iv} are  
$2.1\times10^{-11}$ cm$^3$~s$^{-1}$ and $2.6\times 10^{-11}$ cm$^3$~s$^{-1}$, 
respectively (Ferland et al. 2009; Badnell 2006)\footnote{These recombination rates 
are about a factor of 5 larger than those of Arnaud \& Rothebflugh (1985), adopted 
in literature of most absorption line studies.}. The ionization of a recombining gas will have 
a memory of the previous ionizing continuum for about a recombination time scale, 
i.e., the ionization of gas is connected with an average continuum over such a time 
scale. Continuum fluctuations 
on time scales shorter than this would not cause significant variations in absorption lines, 
rather, they would smear the correlation. Thus, we can use the interval between the two 
observations as an upper limit of the recombination time. The distribution of 
observational intervals for all pairs in the sample is
shown in Figure \ref{interval}. Note that we do not count the minimum observed interval for 
significant absorption line variation because we use the highest $SNR$ spectrum of an object 
as a reference. 

Since absorption line variability is caused by ionizing continuum variations, 
constraints on the gas density can be imposed using the absorption 
line variability time scales (Hamann et al. 1997). Filiz Ak et al (
2013) showed that the fraction of BAL variations increases with an increasing time 
interval between the two observations, with the shortest detected changes in less 
than 10 days for C~{\sc iv} (see also Capellupo et al. 2013). With this time 
scale for C~{\sc iv}, one can set a lower limit on the electron density to 
be $n_e\geq (\alpha_r t)^{-1}\simeq (4-8)\times 10^4$ cm$^{-3}$. 
For a typical AGN continuum and 
column density $>10^{22}$ cm$^{-2}$, the ionization parameter should be 
larger than 1 (refer to \S \ref{secmodel}), so the size of the absorption line 
region should be less than a few tens' of parsecs. It should be pointed out 
that the lack of variability on shorter time scales does not necessarily mean that 
we are detecting a minimum recombination/ionizing time scale. The power spectrum of 
AGN variability is rather red with a power-law of slope about -2.0 on time scales 
from days to several years (e.g., MacLeod et al. 2012), 
so the lack of short time scale variability may be entirely attributed to very small 
variability amplitudes of the ionizing continuum at such short time scale.   

\subsection{On the Emergence and Disappearance of Absorption Lines}

If gas is dominated by higher ionization species, the emergence/disappearance 
of new C~{\sc iv} BAL troughs as the continuum weakens/brightening can be understood. The 
column density of a specific ion is initially too low to be detected, but when 
the ionizing continuum weakens, the species with the higher ionization recombines 
to raise the correspondent ions to a sufficient enough fraction to be detected. 
This actually has been observed in Si~{\sc iv} lines associated with persistent but 
variable C~{\sc iv} absorption line troughs. We found eight such cases in our 
sample, and two examples are shown in Figure \ref{sivemerge}.
The emergence or disappearance 
of a C~{\sc iv} absorption line component is an analog of such variations of Si~{\sc iv} 
but it occurs at a higher ionization level. Large variations in the ionic column density can be 
attributed to either large amplitude of the continuum variation or to the large 
response of the ionic column density to the continuum variations. 

In order to show that large changes in the ionic column density can be produced by
continuum variations, in the following, we  explore analytically 
the physical parameter ranges that an ionic column density is very sensitive to 
the ionizing continuum variations from photoionization equilibrium analysis. We will take 
C$^{3+}$ as an example. The equilibrium of C$^{3+}$ is maintained by a balance 
between source terms, the recombination of C$^{4+}$ plus photoionization 
of C$^{2+}$, and sink terms, the recombination and photoionization of $C^{3+}$
(Osterbrock \& Ferland 2006): 
Ignoring three-body recombination at these low gas densities, we find
\begin{eqnarray}
source - sink = 0\\
n_e n_{C^{4+}}\alpha_{C^{4+}}+ n_{C^{2+}}\int_{\nu_{C^{2+}}}^{\infty}a_{C^{2+}}(\nu)\frac{J(\nu)}{h\nu} d\nu-n_{C^{3+}}\int_{\nu_{C^{3+}}}^{\infty} a_{C^{3+}}(\nu)\frac{J(\nu)}{h\nu} d\nu-n_e n_{C^{3+}}\alpha_{C^{3+}}=0     
\end{eqnarray}
If carbon is dominated by $C^{4+}$,  the terms of recombination of $C^{3+}$ and 
photoionization of $C^{2+}$ will be relatively small in comparison with the other two terms. 
So we can rewrite the equation approximately as
\begin{equation}
n_e n_{C^{4+}}\alpha_{C^{4+}} \simeq n_{C^{3+}} \int_{\nu_{C^{3+}}}^{\infty} a_{C^{3+}}(\nu)\frac{J(\nu)}{h\nu} d\nu
\end{equation}
If the shape of the ionizing continuum remains the same (in the optically thin case) and if the 
ionization is dominated by higher ionization species, then the number ratio of C$^{3+}$ to 
C$^{4+}$ can be written approximately with, 
\begin{equation}
\frac{n_{C^{3+}}}{n_{C^{4+}}}\propto U^{-1} 
\end{equation} 
If $C^{5+}$ is the dominant species,  then $n_{C^{3+}}\propto U^{-2}$. On the other hand, 
if $C^{6+}$ is the dominant one, $n_{C^{3+}}\propto U^{-3}$. A similar analysis can be 
carried out for other ions, such as $Si^{3+}$ and $N^{4+}$. For an optically thick gas, 
the rapid depletion of higher ionization species near the ionization front 
would make the slope even steeper than in the above optically thin case. 
Photoionization model calculations (refer \S\ref{secmodel}) 
suggest an even steeper slope that  can reach -5 for Si~{\sc iv}, C~{\sc iv} and N~{\sc v} for some 
extreme parameters. Thus, the column density variations can be large even 
if the continuum variation is only modest, providing that the ionization is high 
and the total column density is sufficiently large. One should note that if there is 
shielding gas, the situation would occur because shielding gas is more 
transparent to the ionizing continuum at higher photon energies, leading to the 
ionizing continuum softening as the continuum is extinguished. 

If above interpretation is correct, an interesting implication is that there are 
persistent high ionization outflows, which may not be detectable in the UV. 
But when the continuum weakens considerably, the gas ionization lowers to a level 
so that an appreciable fraction of ions are in the correct ionization state, so one can 
detect UV absorption lines. These outflows may  already have been detected in 
 X-ray spectra. Gallagher et al. (2005) detected strong X-ray absorption 
(column density of $\leqslant 10^{22}$ cm$^{-2}$) in quasars with large C~{\sc iv} blueshifted 
emission lines but without absorption lines, and they suspected that the X-ray 
absorbing gas are highly ionized. Such highly-ionized outflows may be an analog 
to the so called ultrafast outflows (UFO) as observed in the X-ray spectra of 
Seyfert galaxies (Tombesi et al. 2012). Note that the typical velocity of UFO 
is 0.1 c, which overlaps with BAL outflows, as do their column densities (Hamann 
et al. 2013). 

\subsection{Photoionization Models\label{secmodel}}

In order to see how ionic column densities change with changes in the ionizing continuum, 
we run a series of photoionization simulations using version c13.03 of Cloudy, last 
described by Ferland et al. (2013). We consider a typical gas density of $10^6$ 
cm$^{-3}$, since gas ionization is not sensitive to density at a given ionization 
parameter $U$.  In practice, we compute  models in $-3\leqslant\log U\leqslant3$ with 
a step $\Delta \log U=0.1$, and $20\leqslant\log N_H\leqslant 24$ with a step $\Delta 
\log N_H=0.2$. Previous observations showed that BAL gas is optically thin to 
the ionizing photons (e.g., Lu et al. 2008; Baskin \& Laor 2013), suggesting that
there is no hydrogen ionization front within outflows. Thus we need only 
to consider a fraction of the parameter space on $\log U$ vs $\log N_H$ 
diagram, which runs diagonally in the $\log U$ vs $\log N_H$ diagram. 

We use the following form of the SED (Mushotzky \& Ferland 1984) to describe the ionizing 
continuum shape
\begin{equation}
f_\nu=A\nu^{-\alpha_{UV}} \exp(-h\nu/kT_{BB})+B\nu^{-\alpha_X}
\end{equation}
where the first term represents the big blue bump in the UV and the second term for the 
power-law X-ray continuum. The parameter $kT_{BB}$ describes the cutoff energy of 
thermal emission from the accretion disk, while $\alpha_{ox}=-\log (f_{2 keV}/f_{2500 \AA})
/\log (\nu_{2 keV}/\nu_{2500\AA})$ characterizes the relative contribution of the power-law 
in X-ray band, i.e., related to the ratio of $A$ and $B$. There are still large uncertainties 
in the ionizing continuum shape of quasars in the far-UV band.  It is inaccessible for low 
redshift quasars due to the Galactic interstellar medium and there are large uncertainties 
in the intergalactic absorption corrections for high redshift quasars. Therefore, we adopt 
the above empirical description rather than HST or FUSE composite spectrum of quasars (Zheng 
et al. 1997; Stevans et al. 2014).
We choose two different $kT_{BB}= 1.5\times 10^5$K and  $2\times 10^5$K
to examine the effect of far-UV continuum, 
and two $\alpha_{ox}=-2.0$, $-1.65$, representing the effect of our ignorance 
of the intrinsic X-ray strength for these quasars (Gallagher et al. 2006; 
Fan et al. 2009) on the final results.  The slope $\alpha_{UV}$ is held fixed at -0.5.

It is worth noting that if there is shielding gas, as proposed by Murray et 
al. (2005) to prevent the outflow being overionized so to keep line driving 
acceleration effective, the ionizing continuum incident on the outflow will 
be more complicated. While large X-ray absorption (equivalent to cold absorption 
of column density $N_H \sim 10^{22}$ to $10^{24}$ cm$^{-2}$ is usually detected in
the X-ray spectrum of BAL QSOs, whether it is highly ionized or not remains uncertain 
due to the weakness of their X-ray emission (Gallagher et al. 2006; Fan et al. 2009). 
It is also controversial whether the X-ray absorber is the same material as the UV 
absorber (Mathur et al. 1998; Wang et al. 2000; Hamann et al. 1998) or it is the 
shielding gas at the base of the outflow. 
Compton thick X-ray absorption has been reported for some low ionization BAL 
QSOs based on the weakness of X-ray emission (e.g., Morabito et al. 2011). Because 
the UV continuum will be subject to very large attenuation due to electron scattering, 
the intrinsic UV luminosity would need to be too high to be correct. However, the conculsion 
is based on the assumption that BAL QSOs have a similar intrinsic X-ray to optical 
luminosity ratio, which need not be true (Fan et al. 2009; Morabito et al. 2014). 
In addition, there is no X-ray absorption in high velocity 
mini-BAL QSOs, suggesting that such shielding gas is not necessary for an outflow to 
be accelerated (Hamann et al. 2013). In fact, if outflows have the high ionization 
parameters discussed here, it is plausible that the X-ray absorption observed in the 
BAL quasars can be formed in the outflow itself.  The recent emergence of both X-ray 
absorption and UV absorption lines in the Seyfert galaxies Mrk 335 and NGC 5548 seems to support 
this scenario (Longinotti et al. 2013; Kaastra et al. 2014). Future simultaneous 
observations of transient BAL events in X-ray and UV can test this. 

Only equilibrium models are considered, i.e., we assume that continuum variations 
are slower than the gas recombination or photoionization timescales. This is likely 
a fair approximation because the ion column density change traces fairly well the 
continuum variations. The requirement that the recombination time of C~{\sc iv}, 
N~{\sc v} and Si~{\sc iv} be shorter than than the shortest variability time scale 
explored in this paper ($\sim$ 0.1 years) converts to a density of 
$n>2\times 10^4$~cm$^{-3}$ (see \S \ref{implication}). 
Non-equilibrium ionization will require time-dependent photo-ionization models 
and knowledge of the continuum light curve. It is beyond the scope of this 
paper, but may be necessary in interpreting high cadence spectroscopic 
monitoring data.   

\begin{figure}
\plottwo{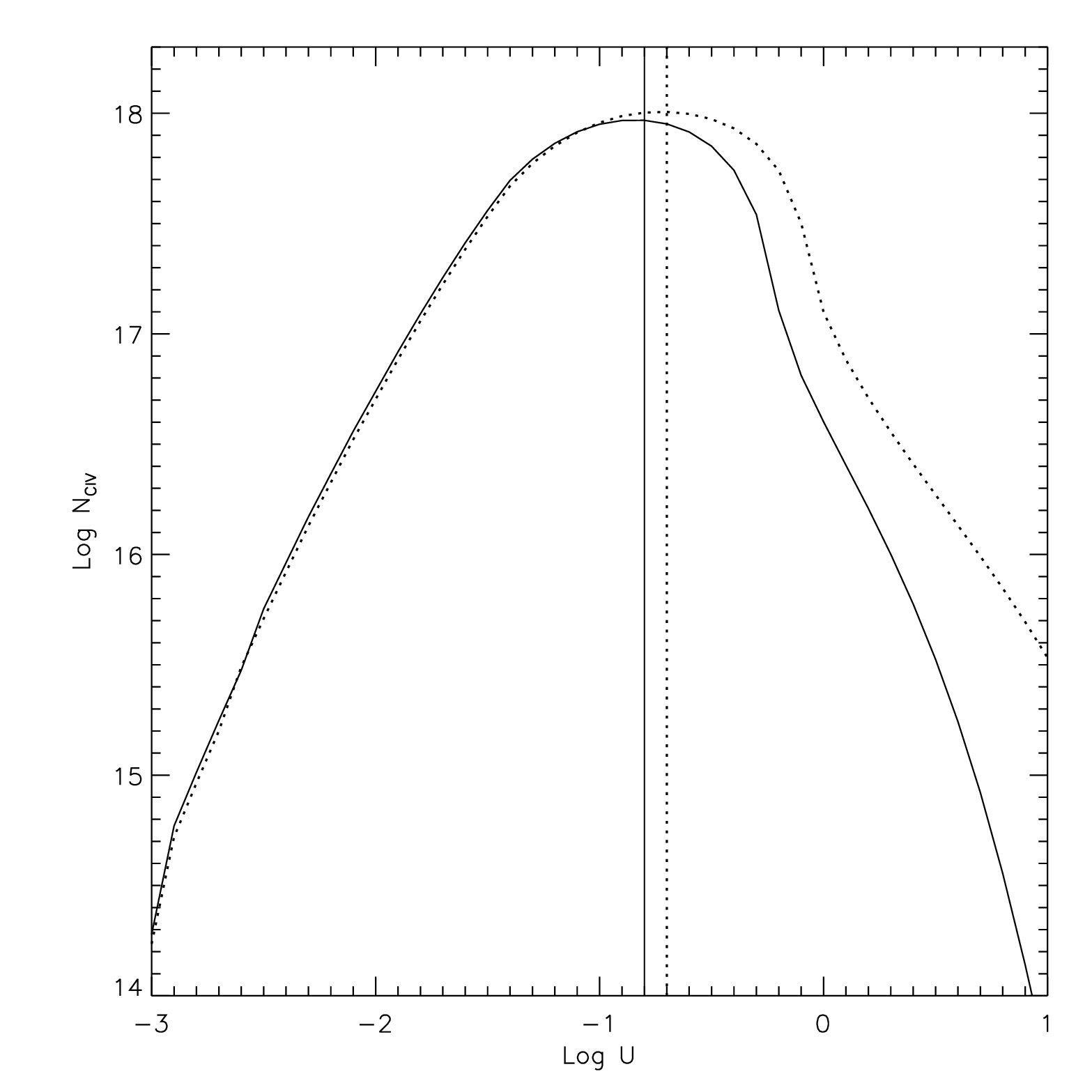}{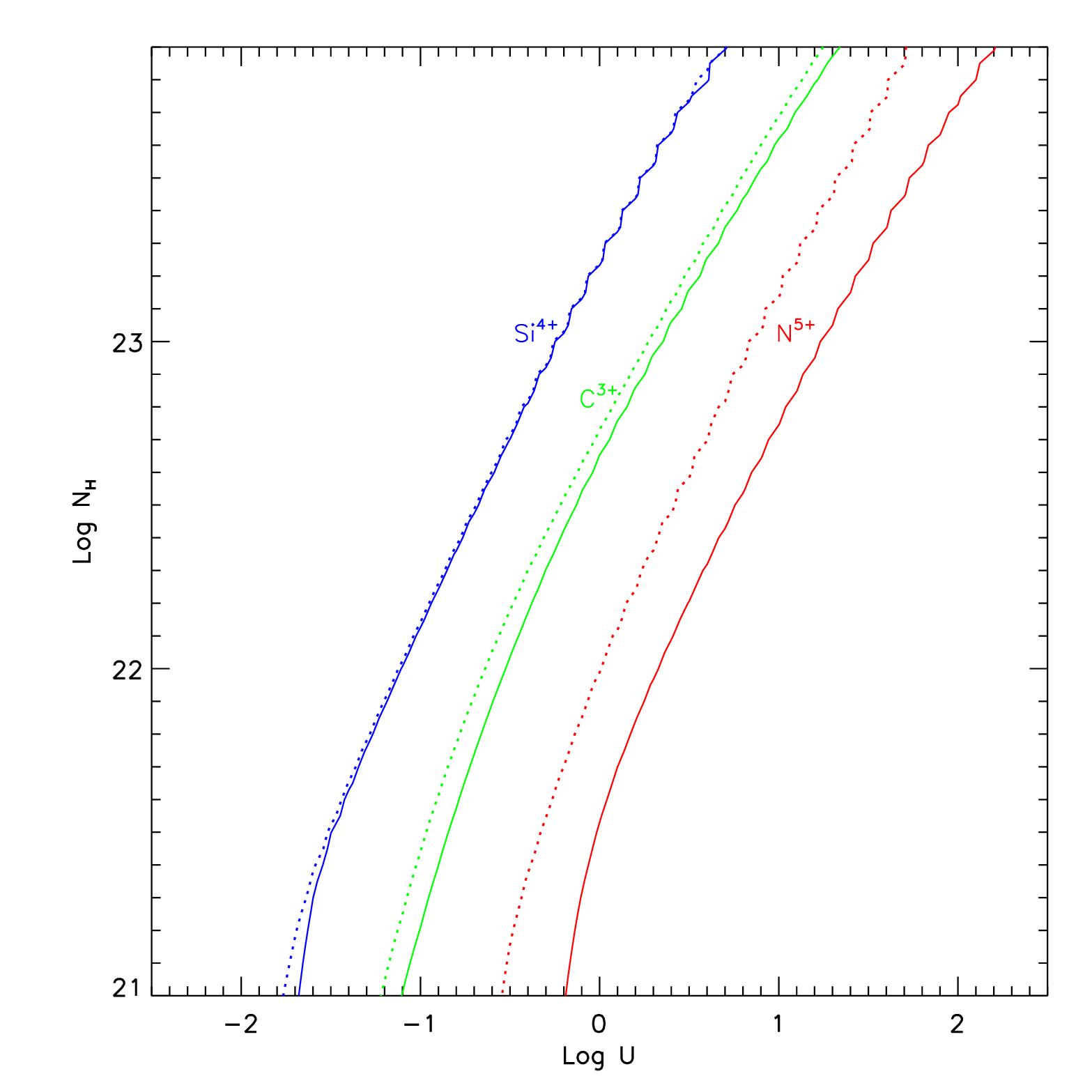}
\caption{
Left panel: The C~{\sc iv} column density as a function of ionization parameter $\log U$ for the 
gas column density $N_H=10^{21.72}$ cm$^{-2}$. The vertical line marks the $\log U$ at 
the maximum fraction of C~{\sc iv}. 
Right panel: The critical boundary that separates positive (to left side) and negative (
to right side) response of ionic column densities to the variations of ionization 
parameters in the $\log U vs \log N_H$ plane. Different colors represent different 
ions: blue line for Si$^{3+}$, red line for N$^{4+}$, and green line for C$^{3+}$. 
For clarity, only two models for $T_{BB}=2\times 10^5$K were plotted (solid line 
for $\alpha_{ox}=2.0$ and dotted line for $\alpha_{ox}=1.65$, see text for more 
detail).
\label{models}}
\end{figure}

\begin{figure}
\plotone{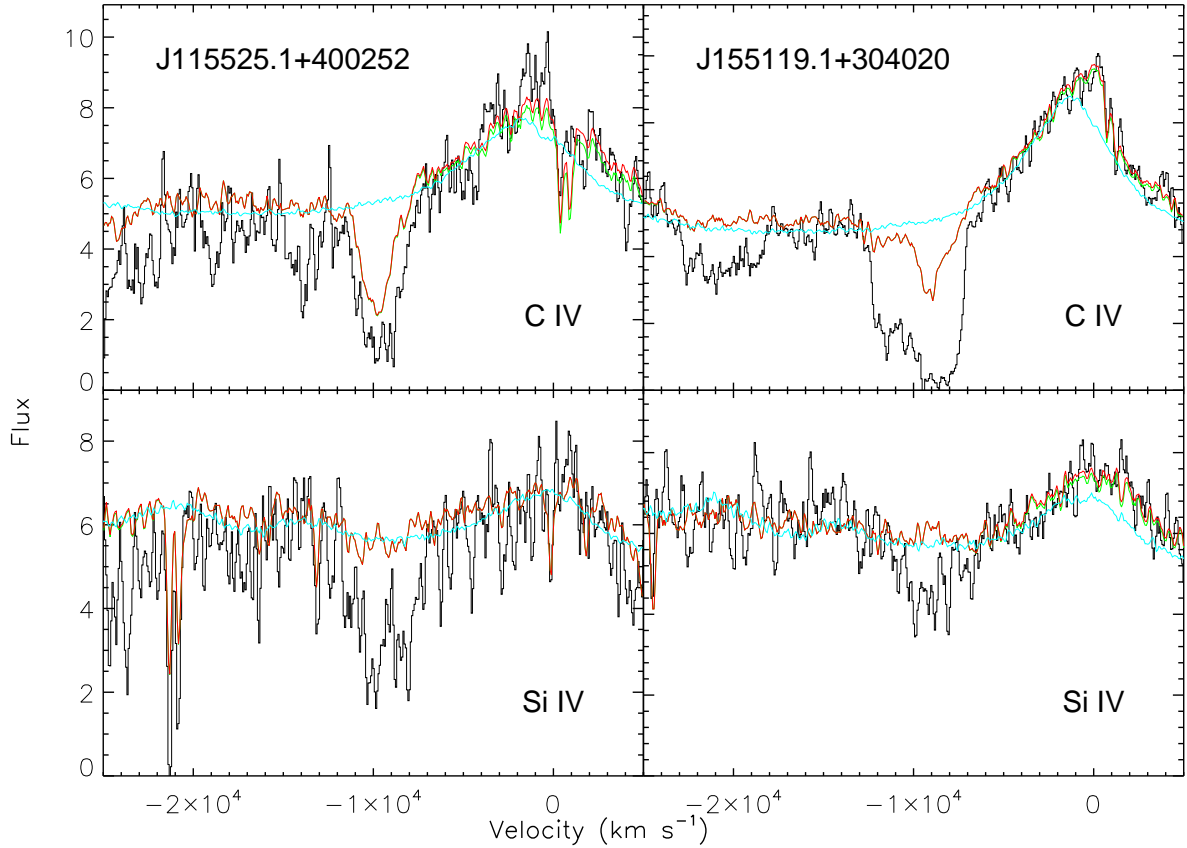}
\caption{Two examples of emergence or disappearance of Si~{\sc iv} absorption lines. 
The SDSS spectra at one epoch and the scaled reference spectrum at another epoch are 
shown in black and green, respectively, as in Figure \ref{specfit}, while the unabsorbed 
QSO template is displayed in cyan. In the reference spectra (red), C~{\sc iv} absorption 
line is prominent, while Si~{\sc iv} absorption line is not significantly detected.}  
\label{sivemerge}
\end{figure}

\begin{figure}
\plotone{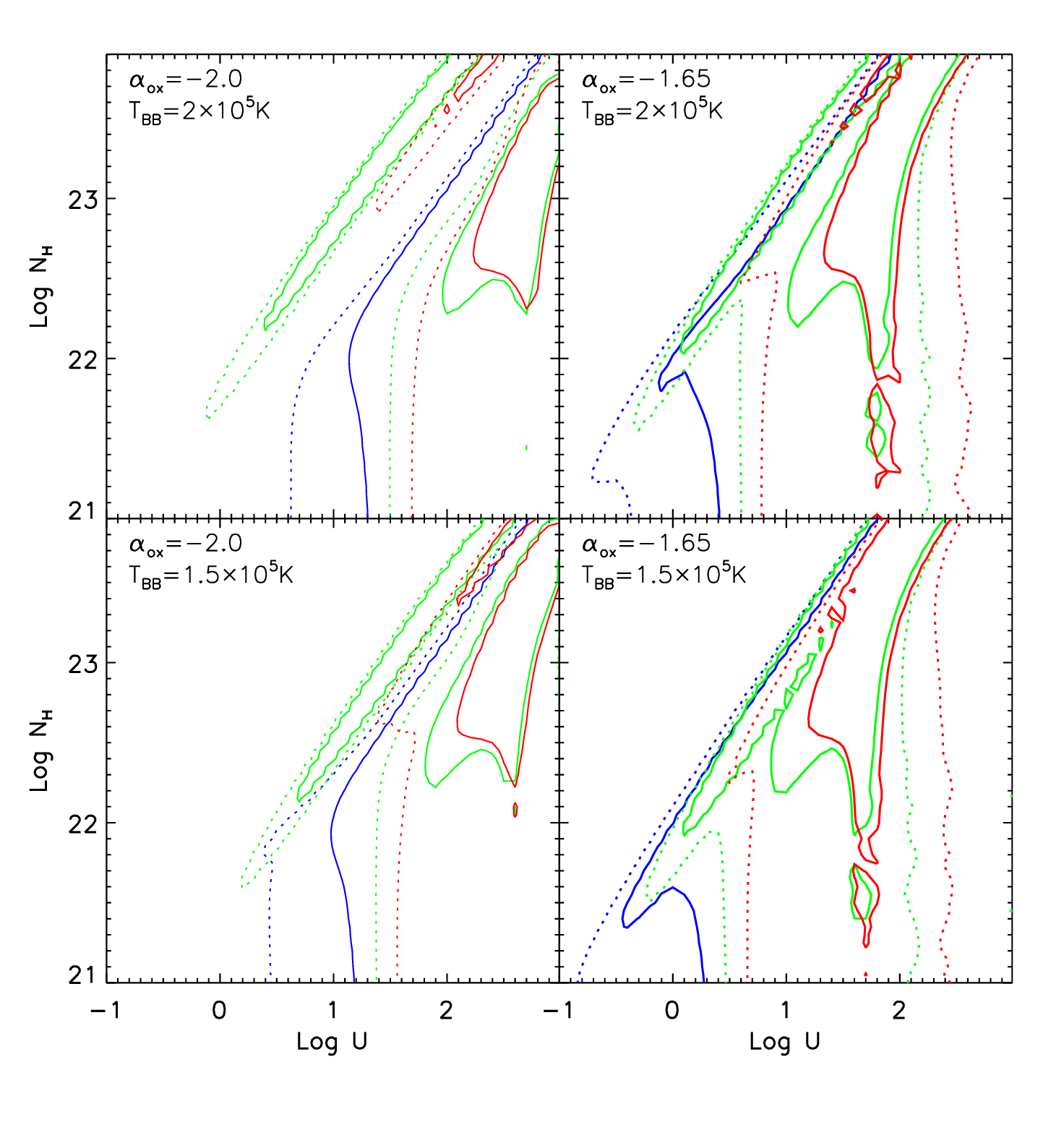}
\caption{The contour for large negative response of Si$^{3+}$ (blue), N$^{4+}$ 
(red) and C$^{3+}$ (green). The dashed line is for $d\log N/d\log U=-3$, and 
the solid line for $d\log N/d\log U=-5$. The response of Si~{\sc iv} increases 
smootly with increasing ionization parameters, while those of C~{\sc iv} 
and N~{\sc v} show complex pattern on the diagram (see text for reasons). The 
shape parameters of the ionizing continuum are marked on the left corner of each panel. For 
$\alpha_{ox}=-1.65$, at $\log U>2$, the gas is too highly ionized to produce 
significant response of either C~{\sc iv} or N~{\sc v}.  
 \label{response}}
\end{figure}

We show the parameter regimes where C~{\sc iv}, Si~{\sc iv} and N~{\sc v} 
response negatively to an increasing ionization parameter in Figure \ref{models}. 
They are located on the lower right side of the critical curves in the Figure. 
As expected, the critical curves of Si~{\sc iv} and C~{\sc iv} do not depend 
much on the ionizing continuum shape, while that of N~{\sc v} changes 
substantially with the two SEDs adopted here. That is because the two SEDs 
result in a large difference in the N~{\sc v} ionizing photons at a given 
$\log U$. Apparently, our results suggest that the ionization of the gas is 
high and most populated species of nitrogen and carbon are in an ionization 
stage higher than Li-like ions. A critical examination of variability of 
higher ionization species such as O~{\sc vi} and Ne~{\sc viii} would be 
important to constrain how high the ionization of gas might be. We also 
show the contours of response of ion column densities to variations of the 
ionization parameter  ($d\log N/d\log U$) in Figure \ref{response}. Apparently, 
at large ionization parameters and large column densities, the response is 
very large with a slope as steep as -5 for C~{\sc iv}, N~{\sc v} and 
Si~{\sc iv} in some zones of the diagram. One interesting feature 
in the figure is that C~{\sc iv} and N~{\sc v} have an additional narrow 
diagonal stripe zone of slope $\leq -5$ to the left such border of the 
Si~{\sc iv}, especially obvious when X-rays are weak. We have checked 
and find that these regions correspond to the formation of an ionization 
front of C~{\sc v} or N~{\sc vi}, on which the ion column density is 
very sensitive to change of ionization parameter. In these parameter 
regimes, a relatively small variation in the continuum luminosity may 
cause a large change in the optical depth of the absorption line, so only moderate 
continuum variation are required to explain the observed disappearance or 
emergence of an absorption line component for a specific ion. The regimes 
are highly overlaped for N~{\sc v} and C~{\sc iv} in the higher ionization 
parameter zone, suggesting that emergence or disappearance of N~{\sc v} 
and C~{\sc iv} may occur at the same time. A thorough analysis of this 
will be given in the future after properly modeling the spectrophotometric 
calibration uncertainty. 

\section{Conclusion} 

We analyze the variability of BALs and mini-BALs and their correlations with 
those of the continuum and emission lines for a sample of 452 quasars in the 
redshift range $2.2<z\leqslant4.7$ and with multi-epoch SDSS spectroscopic 
observations. Variations among different components of C~{\sc iv}, N~{\sc v} 
and Si~{\sc iv} absorption lines, or the same component of different lines, 
are highly coordinated. These conclusions are consistent with previous studies of 
C~{\sc iv}, Si~{\sc iv}, and extend to N~{\sc v} (Hamann et al. 2011; 
Capellupo et al. 2012; Filiz Ak et al. 2013). We find that variations of 
these absorption lines are also highly synchronized with those of the continuum 
and emission lines. The absorption lines weaken/strengthen statistically when 
the continuum brightens/dims. The uncertainties in the continuum flux calibration 
prevent us from assessing the detailed fraction of quasar absorption 
lines where the absorption line variability does not follow continuum variations.  
We also found 50 cases of the emergence or disappearance 
of an absorption line component that accompanies changes in the continuum, 
We interpret these results as indicating that variations in the ionizing continuum 
are the main driver for the absorption line variability 
, and that the dominate species are ions higher than the 
observed Li-like ions for carbon, nitrogen. For a reasonable ionizing continuum 
and gas column density, these constraints imply an ionization parameter of 
$\log U\geqslant 0$. In the case of the disappearance and emergence of BAL 
components, the ionization parameters should be even higher if continuum 
luminosity does not change by a large factor during the period, indicating 
presence of persistent highly ionized outflows even when UV absorption lines 
have disappeared. This can be further tested with future X-ray observations. 
We cannot rule out the possibility that gas moving in or out of line of sight 
may account for the variation of BALs, including their emergence and disappearance, 
in some quasars.
    
\acknowledgements

We thank the referee for useful comments. We acknowledge the financial support 
by the Strategic Priority Research
Program “The Emergence of Cosmological Structures” of the Chinese Academy 
of Sciences (XDB09000000), NSFC through (NSFC-11233002, NSFC-11421303, 
U1431229) and National Basic Research Program of China (grant No. 2015CB857005). 
GJF is grateful to the Leverhulme Trust for support via the award of a Visiting 
Professorship at Queen’s University Belfast (VP1-2012-025). GJF acknowledges 
support by NSF (1108928, 1109061, and 1412155), NASA (10-ATP10-0053, 10-ADAP10-0073, 
NNX12AH73G, and ATP13-0153), and STScI (HST-AR-13245, GO-12560, HST-GO-12309, 
GO-13310.002-A, and HST-AR-13914). Funding for 
SDSS-III has been provided 
by the Alfred P. Sloan Foundation, the Participating Institutions, the National 
Science Foundation, and the U.S. Department of Energy Office of Science. The 
SDSS-III web site is http://www.sdss3.org/. SDSS-III is managed by the 
Astrophysical Research Consortium for the Participating Institutions of the 
SDSS-III Collaboration.  



\newpage

\section*{Appendix A: Effects of Spectrophotometric Calibration Uncertainties on 
the Sign of Continuum Variations of Variable BAL QSOs}
 
The fiber positions of BOSS quasar targets were purposefully offset in order to optimize 
the throughput of light at 4000\AA, while the standard stars used for flux calibration are 
positioned for 5400\AA. This results in a large uncertainty in the flux calibration of 
quasar spectra in the BOSS survey (Dawson et al. 2013). In this appendix, we will estimate 
the effect of 
spectrophotometric calibration uncertainties on the statistical properties of the sign 
of the continuum variation defined in \S 3 using stars that were accidentally observed as 
quasars in the SDSS legacy and SDSS BOSS and $r<20.5$. We assume these stellar spectra 
have similar calibration uncertainties to the spectra of BOSS quasar targets (See Dawson 
et al. 2013). 

We extract the photometric and spectroscopic synthetic magnitudes of these stars from the
SDSS DR10 database. The query resulted in 98,079 stars. We remove M stars, which usually 
have faint magnitudes and very red colors, and so differ from those of most quasars in the 
sample. This leaves a sample of 79,237 stars, including 37,487 from SDSS legacy program. 
We calculate the differences between photometric and spectrosysnthesis magnitudes  
for stars observed in legacy and BOSS programs. The difference is caused by a combination 
of photon noise, imperfect background subtraction, possible variability of some of these 
stars, and the systematic flux calibration uncertainty. The first two terms are relatively 
small in comparison with the observed difference because the typical uncertainty in magnitude 
given by the SDSS pipeline is 0.029 mag in $g$. Most stars should be stable at a level of 
one percent. Neglecting the first three terms, we attribute the difference mostly to the 
flux calibration error, and thus, give a conservative estimate of the latter. 

\begin{figure}
\plotone{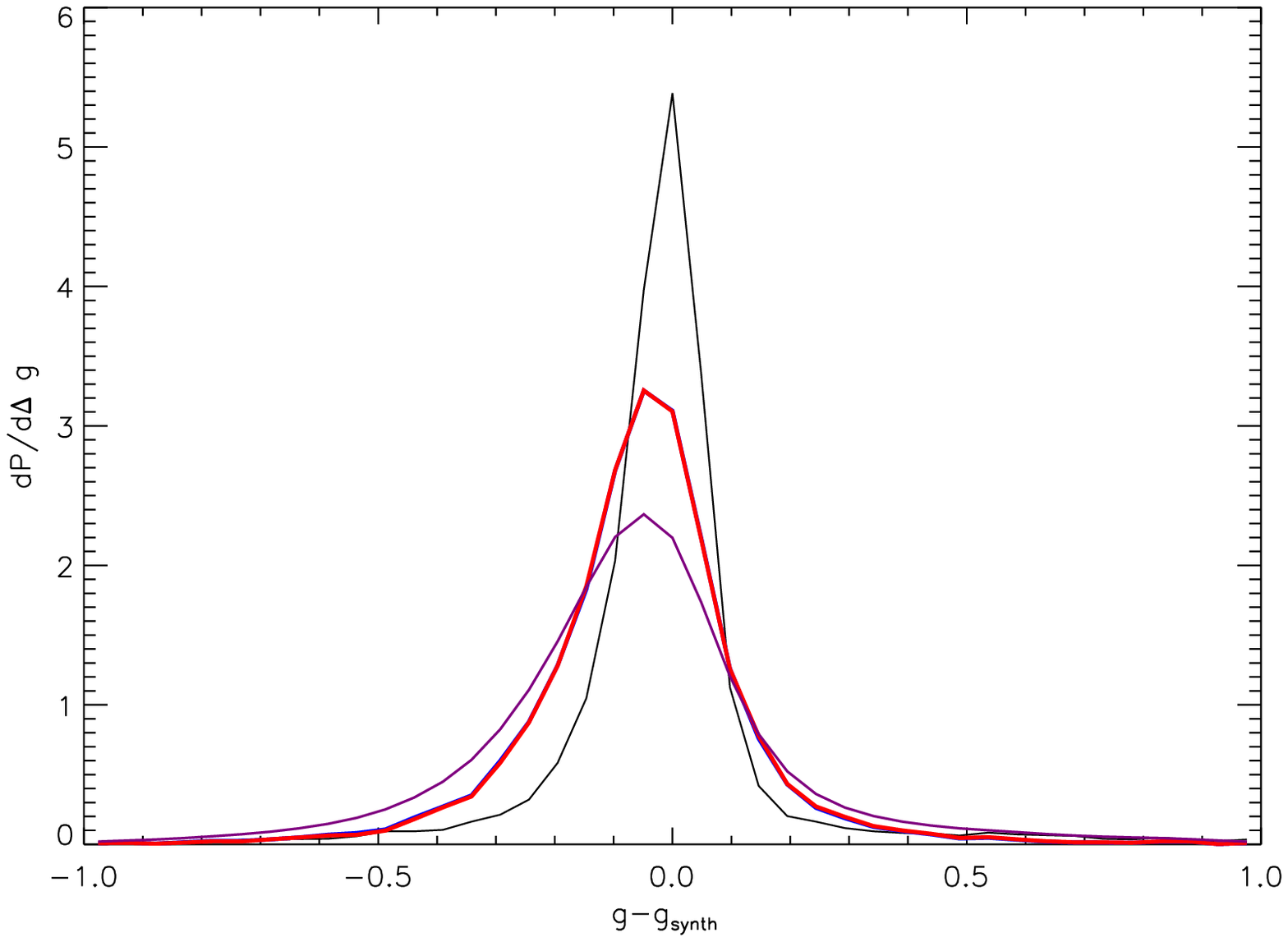}
\caption{The distribution of difference magnitudes in the photometric psf magnitude and the 
spectroscopic synthetic magnitudes for stars mis-targeted as quasars in the SDSS legacy and 
boss spectroscopic survey programs. Spectroscopic targets in legacy program are plotted as a 
black line, while those in boss program as a blue line. For comparison, the average two-epoch 
magnitude difference spread function (see text for detail) for variable BAL QSO sample is 
also shown as the red line.  
\label{magdiff} }
\end{figure}

We plot the distributions of the difference magnitude (DDMs) in the $g$-band ($g-g_{synth}$) 
for stars observed in BOSS and legacy programs, separately, in Figure \ref{magdiff}. There 
is a small (-0.05 mag) systematic 
offset in the peak of the distribution for BOSS spectra but 
no such offset in legacy spectra. In either case, the distribution is not symmetric, but 
rather skewed to the negative, i.e., a fraction of stars appear fainter in SDSS spectrum 
than they should be. Also the difference distribution for BOSS spectra is much broader 
than that for legacy spectra, suggesting large calibration errors. When splitting stars 
into different spectroscopic types, we see similar distributions for types B, A, G, F and 
K. Type O stars have a significantly broader wing than other types for unknown reasons. Because of this 
we do not try to do any color related corrections. We also check the distributions in other 
bands, and found that they are significantly broader than in the $g$ band. So we will mainly 
use the $g$-band. This approximately corresponds to a rest frame wavelength of 1400\AA 
for majority of our quasars. 

Next, we estimate the approximate average error distribution of the differential synthetic 
magnitudes of quasar spectra obtained at two different epochs (DDSM for short) using 
the DDMs described above. We convolve the DDM of one spectrum with that of another assuming that
their errors are 
uncorrelated to obtain the DDSM of a specific spectrum pair. Since the spectra taken in 
legacy and BOSS have different DDMs, we obtain four different DDSMs for combinations: 
(legacy, legacy), (legacy, BOSS), (BOSS, legacy) and (BOSS, BOSS). The systematic offset 
is subtracted from the DDM for the BOSS spectra. We make a average of the DDSM (Figure 
\ref{magdiff}), weighted with the number of quasar pairs of our variable BAL quasars 
in each combination. The distribution is fairly broad, suggesting that it is impossible 
to quantitatively estimate the variations of an individual spectrum pair. Note that this 
treatment is only a conservative approximation since quasars in ancillary programs may 
not be observed in the mode of fiber position offset and it was shown that the flux calibration 
error is smaller in the latter case (Dawson et al. 2013). 

\begin{figure}
\plotone{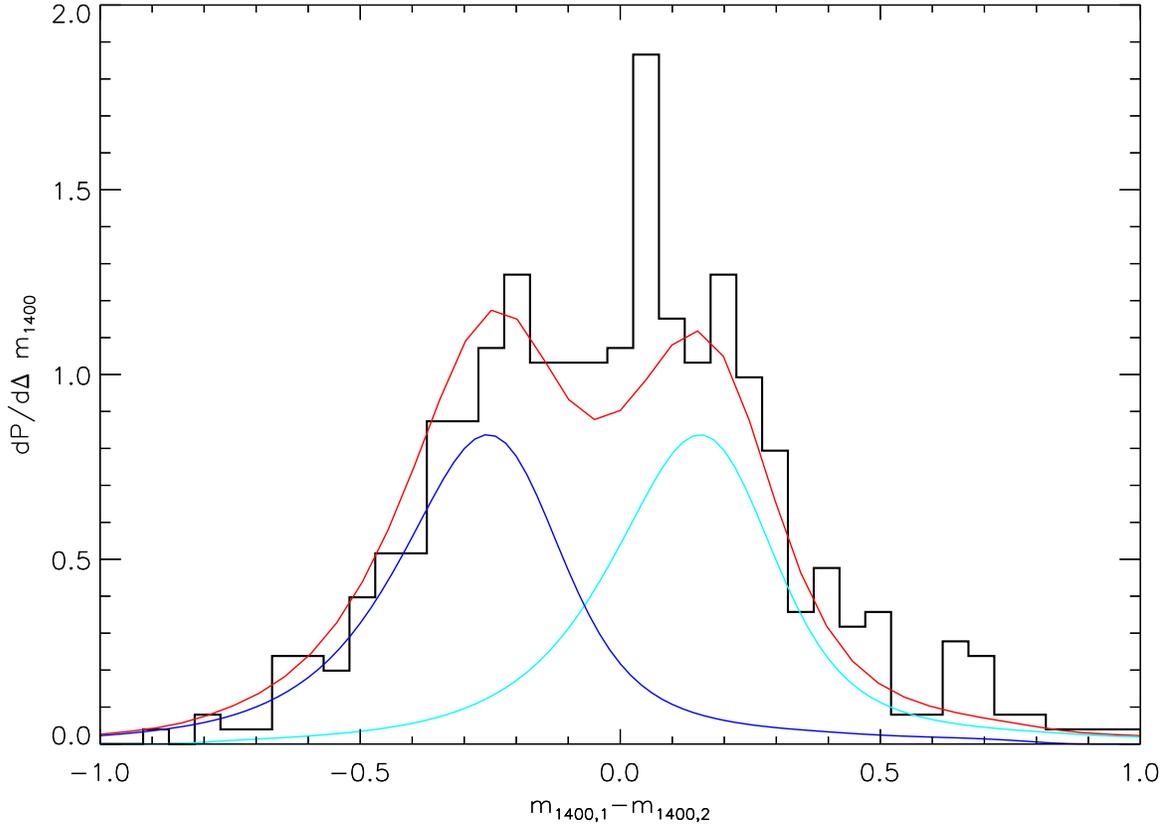}
\caption{The distribution of the difference of spectral synthetic magnitudes at two epochs 
of variable BAL QSO sample. The red line represents the best fitted double gaussian model convolved 
with the two epoch magnitude difference spread function in Figure \ref{magdiff}. The blue and cyan 
lines are the convolved positive and negative part of the intrinsic distributions, respectively. 
\label{temd}}
\end{figure}

Finally, we calculate the two-epoch magnitude difference (TEMD) for quasar pairs in Table 
1 by correcting the systematic offset in the spectrophotometric calibration in the BOSS 
survey for quasars. The difference magnitude at 1400\AA~ is calculated by using the double 
power-law scaling factor (Equation 1) in the reference matching procedure (\S 2.2). The 
distribution of TEMD (Figure \ref{temd}) is considerably broader than the average DDSM, 
suggesting that variability information can be extracted statistically. The distribution 
cannot be fitted by a single Gaussian function convolved with DDSM, so we fit it with the 
sum of two Gaussian functions. Figure \ref{temd} shows the best fit and the double Gaussian 
model for the intrinsic distribution.  We take the latter as the true distribution for 
magnitude variations, and examine what fraction of sources were assigned to a wrong sign 
of variations using the definition in this paper after convolving the positive and negative 
part of the distribution with the two-epoch MSF. We find that 13\% of sources were assigned 
opposite signs due to the calibration uncertainties, and 10\% were assigned uncertain signs, 
i.e., magnitude difference between -0.05 and 0.05 mag. Considering that some quasars in 
the ancillary programs were not observed in an offset mode, we likely over-estimate the 
number of pairs with mis-assigned sign.    



\begin{deluxetable}{cc}
\tabletypesize{\scriptsize}
\tablewidth{0pt}
\tablecaption{Basic Statistics of the C IV Variable BAL Sample \label{sample}}
\tablehead{
\colhead{Variable parameter} & \colhead{Number of Sources} }
\startdata
C~{\sc iv} & 452 \\
Si~{\sc iv} & 194 \\
N~{\sc v} & 227 \\
Emission Lines & 396 \\
Continuum     & 421 \\ \hline
\enddata
\medskip
\vfill
\end{deluxetable} 


\begin{deluxetable}{ccccccccccc}
\tabletypesize{\scriptsize}
\tablewidth{0pt}
\tablecaption{List of Variable Absorption Line Components\tablenotemark{1} \label{tab1}}
\tablehead{
\colhead{Name\tablenotemark{2}} & \colhead{plate} & \colhead{mjd} &\colhead{fiberid} & \colhead{Velocity range} & 
\multicolumn{5}{c}{Variability sign relative to reference\tablenotemark{3}} \\ 
\cline{6-10} \\
\colhead{} & \colhead{} & \colhead{} & \colhead{} & \colhead{km~s$^{-1}$} &
\colhead{C IV} & \colhead{Si IV} & \colhead{N V} & \colhead{continuum} & \colhead{emission line} } 
\startdata
J000330.18$+$000813.2&4217&55478&532& reference &$--$&$--$&$--$&$--$&$--$&\\
 &686&52519&356&$-23900\sim-12400$&$-1$&$-1$&$-1$&$1$&$-1$\\
J000951.17$+$092710.5&4534&55863&968& reference &$--$&$--$&$--$&$--$&$--$&\\
 &5648&55923&386&$-25700\sim-19200$&$1$&$0$&$0$&$-1$&$1$\\
 &5648&55923&386&$-14500\sim-11100$&$1$&$0$&$1$&$-1$&$1$\\
J001130.55$+$005550.7&4217&55478&948& reference &$--$&$--$&$--$&$--$&$--$&\\
 &389&51795&339&$-4900\sim-700$&$1$&$0$&$0$&$1$&$0$\\
 &686&52519&603&$-4900\sim-700$&$1$&$0$&$0$&$-1$&$1$\\
 &687&52518&339&$-4900\sim-700$&$1$&$1$&$0$&$0$&$1$\\
J001818.70$+$002709.1&4218&55479&972& reference &$--$&$--$&$--$&$--$&$--$&\\
 &1491&52996&589&$-16500\sim-7100$&$1$&$0$&$1$&$1$&$-1$\\
J002146.71$-$004847.9&4219&55480&216& reference &$--$&$--$&$--$&$--$&$--$&\\
 &390&51816&161&$-22300\sim-19900$&$1$&$0$&$0$&$-1$&$1$\\
J003135.57$+$003421.2*&3587&55182&570& reference &$--$&$--$&$--$&$--$&$--$&\\
 &689&52262&502&$-19800\sim-7900$&$-1$&$-1$&$--$&$1$&$-1$\\
J004022.40$+$005939.6&4222&55444&710& reference &$--$&$--$&$--$&$--$&$--$&\\
 &690&52261&563&$-10200\sim-3300$&$1$&$0$&$0$&$0$&$1$\\
\enddata
\medskip
\vfill
{\normalsize $^1$ A complete version of this table can be obtained online. } \\
{\normalsize $^2$ Symbol '*' marks an object with appearance or disappearance of 
one or more absorption line components with respect to the reference spectrum.}\\
{\normalsize $^3$ The value 1 represents that case either equivalent width 
of an absorption line or emission line strengthens or continuum brightens in 
comparison with the reference spectrum, and -1 for an absorption or emission 
line weakens or continuum dims, while 0 for insignificant variation, -- for no 
available data.} 
\end{deluxetable} 

\end{document}